\documentclass[10pt,twocolumn,superscriptaddress,amsmath,amssymb,aps,prb]{revtex4-2}

\pdfoutput=1

\usepackage[utf8]{inputenc}
\usepackage[english]{babel}
\usepackage[colorlinks=true,linkcolor=blue,citecolor=blue]{hyperref}
\usepackage{xcolor}
\usepackage{braket}
\usepackage{graphicx}
\usepackage[caption=false,margin=0pt]{subfig}
\usepackage{ulem}
\usepackage{siunitx}
\usepackage{bibunits}

\begin{document}

\renewcommand{\vec}{\mathbf}
\renewcommand{\figurename}{FIG.}

\title{Microscopic theory of spin-relaxation of a single Fe adatom coupled to substrate vibrations}

\makeatletter
\let\thetitle\@title
\makeatother

\author{Haritz Garai-Marin}
\affiliation{Physics Department, University of the Basque Country UPV/EHU, 48080 Bilbao, Basque Country, Spain}
\affiliation{Donostia International Physics Center (DIPC), Paseo Manuel de Lardizabal 4, 20018 Donostia-San Sebasti{\'a}n, Spain}

\author{Manuel dos Santos Dias}
\affiliation{Peter Gr\"unberg Institut and Institute for Advanced Simulation, Forschungszentrum J\"ulich $\&$ JARA,
52425 Jülich, Germany}
\affiliation{Faculty of Physics, University of Duisburg-Essen $\&$ CENIDE, 47053 Duisburg, Germany}
\affiliation{Scientific Computing Department, STFC Daresbury Laboratory, Warrington WA4 4AD, United Kingdom}

\author{Samir Lounis}
\affiliation{Peter Gr\"unberg Institut and Institute for Advanced Simulation, Forschungszentrum J\"ulich $\&$ JARA,
52425 Jülich, Germany}
\affiliation{Faculty of Physics, University of Duisburg-Essen $\&$ CENIDE, 47053 Duisburg, Germany}

\author{Julen Ibañez-Azpiroz}
\affiliation{Centro de F{\'i}sica de Materiales, Universidad del Pa{\'i}s Vasco UPV/EHU, 20018 San Sebasti{\'a}n, Spain}
\affiliation{IKERBASQUE Basque Foundation for Science, 48013 Bilbao, Spain}

\author{Asier Eiguren}
\affiliation{Physics Department, University of the Basque Country UPV/EHU, 48080 Bilbao, Basque Country, Spain}
\affiliation{Donostia International Physics Center (DIPC), Paseo Manuel de Lardizabal 4, 20018 Donostia-San Sebasti{\'a}n, Spain}
\affiliation{EHU Quantum Center, University of the Basque Country UPV/EHU}

\date{\today}

\begin{abstract}

Understanding the spin-relaxation mechanism of single adatoms 
is an essential step towards creating atomic magnetic memory bits or even qubits. 
Here we present an essentially parameter-free theory
by combining \textit{ab-initio} electronic and vibrational properties
with the many-body nature of atomic states. 
Our calculations account for the millisecond
spin lifetime measured recently 
on Fe adatoms on MgO/Ag(100) 
and reproduce the dependence on the number of decoupling layers and 
the external magnetic field. 
We show how the atomic interaction with the environment
should be tuned in order to enhance the magnetic stability, and 
propose a clear fingerprint for experimentally detecting 
a localized spin-phonon excitation. 
\end{abstract}

\maketitle

The development of cutting-edge experimental techniques for manipulating and probing nano-structures has allowed the study of a large variety of quantum effects at the atomic scale. In particular, magnetic single adatoms offer an exceptional scenario to explore spin excitations~\cite{Heinrich2004,Loth2010a,Ternes2015}, magnetic interactions~\cite{Otte2008,Meier2008,Bouaziz2020} or spin relaxation and decoherence~\cite{Delgado2017,Donati2016,Paul2017,Natterer2018}, among other phenomena. The control of the underlying physics has enabled the optimization of their magnetic properties with potential applications in mind, such as high-density storage devices or quantum computing~\cite{Miyamachi2013,Rau2014,Loth2010,Natterer2017}. Understanding how adatoms interact with their environment is also essential to continue designing systems with longer spin relaxation times as well as improving their control and manipulation techniques.

The upsurge of inelastic electron tunneling spectroscopy~\cite{Heinrich2004} has opened the way to access relevant information from magnetic excited states~\cite{Fernandez-Rossier2009,Lorente2009}. In addition, electron spin resonance combined with scanning tunneling spectroscopy has provided a boost on energy and spatial resolution, allowing remarkable achievements~\cite{Baumann2015b,Natterer2017,Willke2018,Yang2018,Yang2019,Yang2019a}. Several theoretical models have been proposed to capture the essential physics~\cite{ReinaGalvez2019,Seifert2020,Delgado2021}, but the physical origin of the spin transitions is still an open question. In the case of the long-living magnetic states found in adatoms such as Ho or Fe on MgO/Ag(100)~\cite{Donati2016,Natterer2018,Paul2017}, interactions with the environment are believed to destabilize the magnetic moments by inducing transitions between the different spin states. However, the difficulty of performing first-principles calculations of spin lifetimes and transitions rates complicates verifying the origin of the physical mechanisms involved in  experimental measurements.

While the effect of electronic interactions on adatom properties has been widely studied using first-principles schemes~\cite{Lorente2009,Lounis2010,Khajetoorians2011,Yang2011,Lounis2015,Ferron2015a,Khajetoorians2016,Ibanez-Azpiroz2016,Hermenau2017,Ibanez-Azpiroz2017,Ibanez-Azpiroz2017a,Wolf2020}, much less attention has been paid to the interaction with substrate vibrations, namely phonons. These are nevertheless believed to play a key role in the many experiments that make use of insulating decoupling layers, such as Cu\textsubscript{2}N or MgO, which reduce the interaction with substrate conduction electrons~\cite{Hirjibehedin2007,Donati2016}. \textit{Ab-initio} calculations of the spin-phonon coupling have a long track record in the field of molecular magnets~\cite{Lunghi2017,Lunghi2017a,Escalera-Moreno2017,Escalera-Moreno2018,Albino2019,Lunghi2019,Lunghi2020,Briganti2021}, but to our knowledge there are only two studies concerning single adatoms on surfaces: a study of the temperature dependent relaxation mechanism of Ho on MgO/Ag(100)~\cite{Donati2020} and an initial characterization of the spin-diagonal electron-phonon interaction of Fe on MgO/Ag(100)~\cite{Garai-Marin2021}.

In this work, we present a methodology that combines first-principles density functional theory (DFT) calculations with an atomic multiplet model for accessing the electron-phonon spin relaxation time, with a concrete application to a single Fe atom on MgO/Ag(100). The DFT calculations allow an accurate characterization of both the spin-phonon coupling as well as the hybridization of the adatom with the substrate, while the atomic model provides the necessary  many-body picture of the atomic electronic states. Our calculations successfully account for the millisecond spin lifetime measured experimentally~\cite{Paul2017} and reproduce two central trends: the dependence on the number of decoupling layers and the external magnetic field. We show that the main relaxation channel involved in the experimental measurements does not correspond to localized vibrations of the adatom, but low-energy acoustic modes of the substrate. We further propose a means of experimentally verifying the relaxation mechanism at play by tuning the external magnetic field to match the energy of the localized phonon mode, where we predict plateau behavior of the relaxation time.

We have obtained the electronic and vibrational properties of the Fe/MgO/Ag(100) system by means of relativistic first-principles DFT calculations as implemented in the SIESTA code \cite{Soler2002,Cuadrado2012}. The vibrational frequencies and the potential induced by the atomic displacements were calculated using the so-called direct method, employing symmetries to reduce computational costs, as explained in Ref.~\cite{Garai-Marin2021}. The unit cell consisted of a $4\times4\times1$ supercell of a MgO/Ag(100) slab, with an iron adatom adsorbed on top of an oxygen site. See Ref.~\footnote[1]{See Supplemental Material at [URL] , which includes Refs.~\cite{Perdew1996,Kleinman1982,Breuer2007}, for a detailed information of the DFT calculations, a list of Stevens operators and the second quantization expression of the basis states, together with the derivation of the rate equation.} for all the details.

\begin{figure}[tb]
\centering
\includegraphics[width=\linewidth]{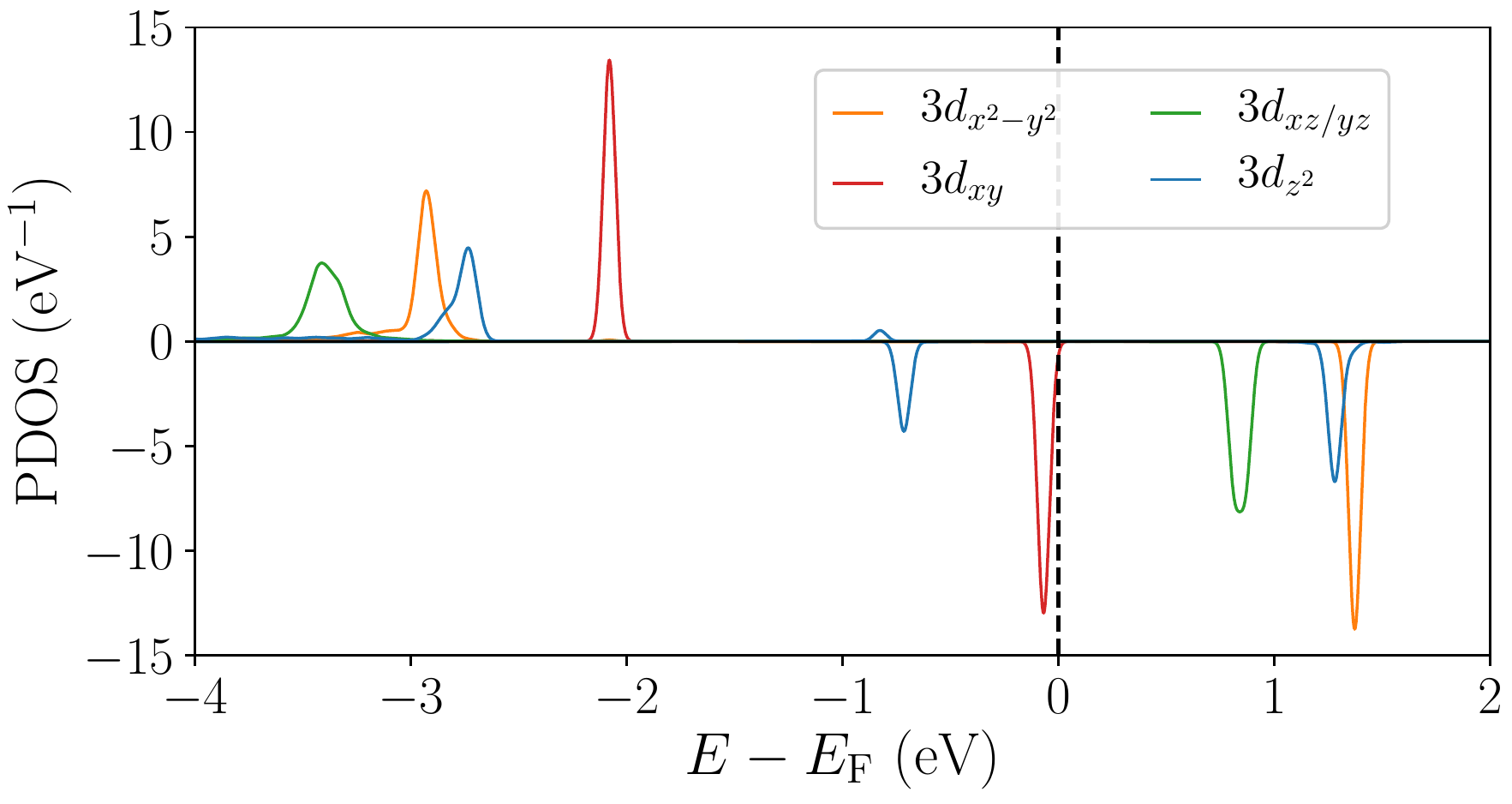}
\caption{\label{fig:pdos} Electron DOS projected onto the Fe $3d$ orbitals when the adatom is placed on 3~ML MgO/Ag(100). Positive and negative values of the PDOS indicate majority and minority spin channels, respectively. The Fermi energy is marked by a vertical dashed line.}
\end{figure}

Our DFT calculations predict a magnetic ground state with a total magnetic moment of 4.06 $\mu_\mathrm{B}$, of which 3.97 $\mu_\mathrm{B}$ is localized on the iron adatom. It is formed by the Fe $3d$ orbitals, of which 5 majority spin and 1 minority spin are occupied. \figurename~\ref{fig:pdos} illustrates this with the spin-resolved projected density of states (PDOS) of the Fe adatom on 3 monolayers (MLs) of MgO. We obtained similar results for different MgO coverages, the major difference being a larger hybridization in the case of 2 MgO layers, mostly in the spin majority $3d_{xz/yz}$, $3d_{x^2-y^2}$ and $3d_{z^2}$ orbitals.

\begin{figure}[tb]
\centering
\includegraphics[width=\linewidth]{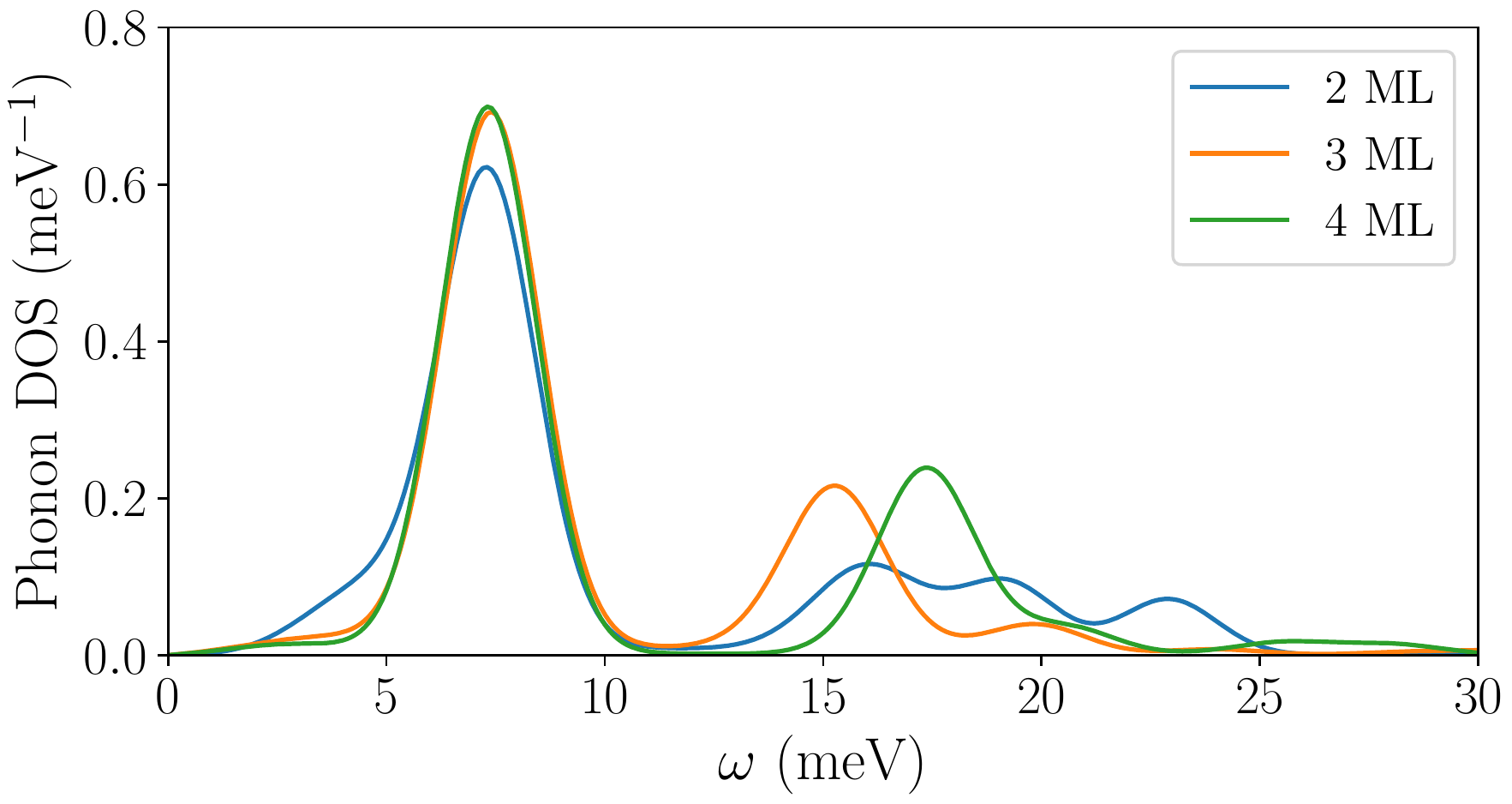}
\caption{\label{fig:ph-dos} Phonon DOS projected onto the Fe adatom deposited on MgO/Ag(100) for 2 to 4 MLs of MgO.
}
\end{figure}

Turning next to the vibrational properties, \figurename~\ref{fig:ph-dos} shows the phonon DOS projected on the iron adatom. There is a doubly-degenerate vibrational mode completely localized in iron at \SI{7.3}{\milli\electronvolt}, which corresponds to an in-plane oscillation of the adatom with respect to the MgO surface. At higher energy, in the 12 to \SI{25}{\milli\electronvolt} range, the multiple modes observed are related to an out-of-plane motion of the adatom. These modes show a more pronounced mixing with the substrate, as can be appreciated from the broadening of the peak.

We now describe the coupling between electrons and phonons. 
In the basis of one-electron atomic orbitals $\ket{\psi_i}$ with quantum numbers $i=\{n, l, m, \sigma\}$ ($n$ is the principal quantum number, $l$ the angular momentum, $m$ its projection and $\sigma$ the spin) and using the second-quantization formalism, 
the electron-phonon coupling to first order in the atomic displacements is described by~\cite{Grimvall1981,Mahan2000,Giustino2017},
\begin{align} \label{eq:el-ph}
H_\mathrm{el.-ph.}
=
\sum_\eta
H_\mathrm{el.-ph.}^\eta
=
\sum_{i,f,\eta}
g_{i, f}^\eta \, c^\dagger_{f} c_{i} \, (b^\dagger_\eta + b_\eta)
.\end{align}
Above, the electron-phonon matrix element $g_{i, f}^\eta$ determines the probability amplitude of an atomic state $\ket{\psi_{i}}$ to be scattered to a state $\ket{\psi_{f}}$ by the emission or absorption of a phonon $\eta$. $c^\dagger$ and $c$ ($b^\dagger$ and $b$) are the electron (phonon) creation and annihilation operators, respectively. In reality, the atomic orbitals hybridize with the substrate and lose spherical symmetry. Even so, it is very useful to keep the associated quantum numbers to be able to perform multiplet calculations with total angular momentum. We take hybridization into account by defining new atomic orbitals $\ket{\tilde{\psi}_i}$ by projection of the DFT spinor wave functions $\ket{\psi_n^\mathrm{DFT}}$,
\begin{align} 
\label{eq:phi0}
\ket{\tilde{\psi}_i} = \sum_n \ket{\psi_n^\mathrm{DFT}} \braket{\psi_n^\mathrm{DFT} | \psi_i},
\end{align}
Normalization factors are absent because we only need the projection of orbitals onto the active DFT-represented subspace.

In this way, the electron-phonon matrix elements between the one-electron atomic orbitals are approximated by
\begin{align} \label{eq:matrix-elements-atomic}
g_{if}^\eta
=
\braket{ \tilde{\psi}_i | \delta V_\eta | \tilde{\psi}_f }
,\end{align}
where $\delta V_\eta$ is the change in the DFT potential caused by a phonon mode $\eta$ (a 2$\times$2 matrix in spin space). Note that by computing the matrix elements in this way we are able to capture the effect of the hybridization between the iron and the substrate, which varies for different MgO coverages. Therefore, the electron-phonon coupling in the substrate area is accounted for in this scheme.

In order to incorporate the many-body nature of the adatom~\cite{Wolf2020}, we now consider a multiplet Hamiltonian in terms of fixed total spin ($\vec{S}$) and total orbital angular momentum ($\vec{L}$) operators. Starting from the $^5D$ term of an isolated iron atom with a $d^6$ configuration, the lowest energy term according to Hund's rules, the crystal field of the substrate can be expanded using Stevens operators $\hat{O}_n^m(\mathbf{L})$~\footnotemark[1]. In the case of Fe on top of an O atom of MgO, it is known that the low energy levels are well described by
\begin{align} \label{eq:Fe-Stevens-Hamiltonian-Baumann}
H
=
B_2^0 \hat{O}_2^0
+
B_4^0 \hat{O}_4^0
+
B_4^4 \hat{O}_4^4
+
\lambda \vec{L} \cdot \vec{S}
+
\mu_\mathrm{B}\left( \vec{L} + 2\vec{S} \right) \cdot \vec{B}
,\end{align}
where $B_2^0 = \SI{-317.43}{\milli\electronvolt}$, $B_4^0 = \SI{-6.58}{\milli\electronvolt}$, $B_4^4 = \SI{-4.36}{\milli\electronvolt}$ and $\lambda = \SI{-12.6}{\milli\electronvolt}$ were obtained from a point charge model~\cite{Baumann2015a}. The first two terms compose the axial crystal field (ACF) and the third one the transverse crystal field (TCF). The fourth term describes the spin-orbit coupling with interaction strength $\lambda$ and the last one accounts for the Zeeman term, with $\mu_\mathrm{B}$ the Bohr magneton and $\vec{B}$ the magnetic field.

\begin{figure}[tb]
\centering
\includegraphics[width=\linewidth]{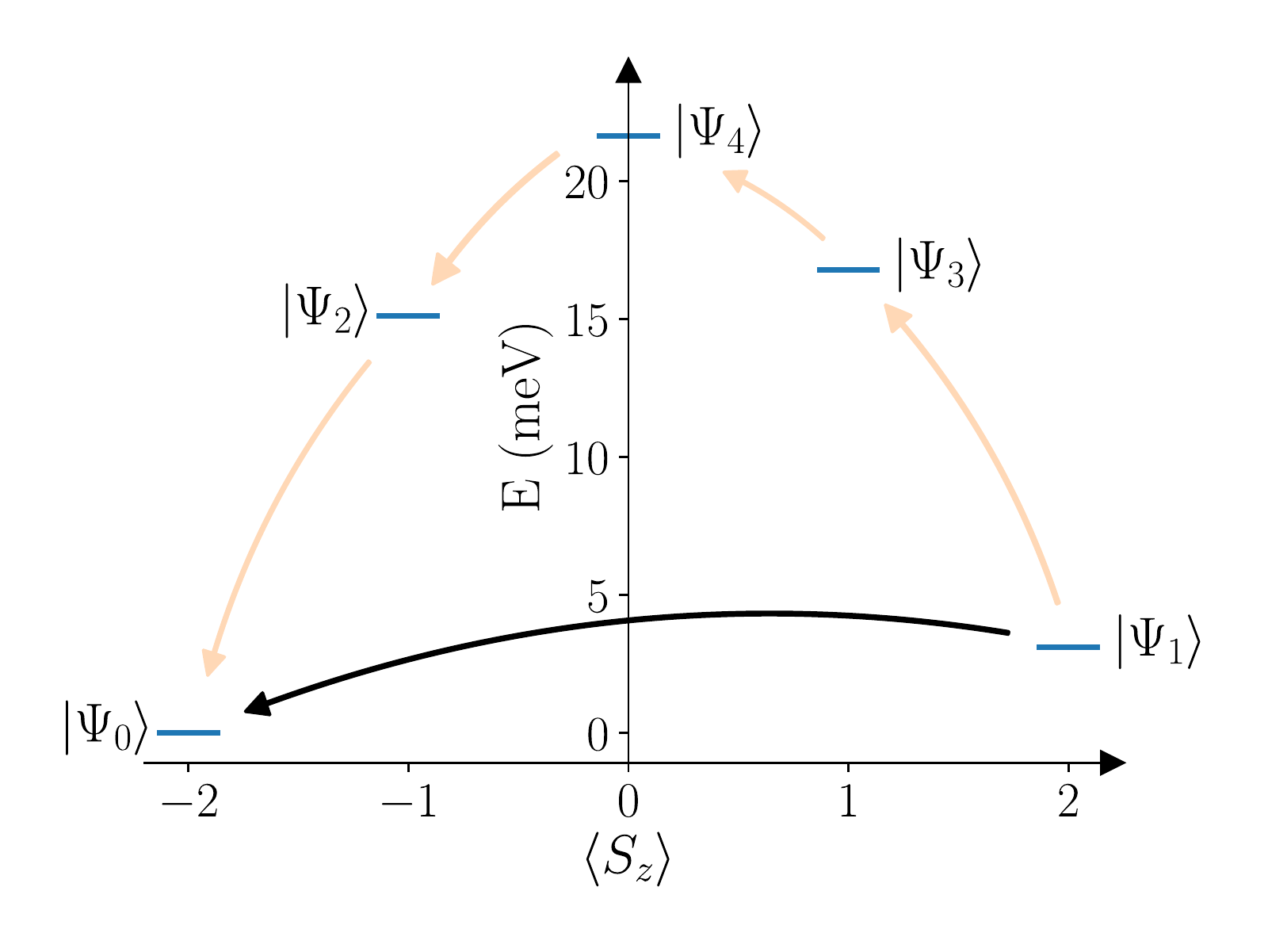}
\caption{Energy level diagram of the 5 lowest energy states of the crystal field Hamiltonian~\eqref{eq:Fe-Stevens-Hamiltonian-Baumann}. The arrows indicate possible pathways for the spin relaxation mechanism, with the first order or direct transition pointed out by the black arrow and the over-barrier transition by the orange arrows.}
\label{fig:energy-diagram}
\end{figure}

Direct diagonalization of the Hamiltonian in Eq.~\eqref{eq:Fe-Stevens-Hamiltonian-Baumann} using the product basis of projections of total spin $M_\mathrm{S}$ and orbital angular momenta $M_\mathrm{L}$, $\{\ket{M_\mathrm{S}, M_\mathrm{L}}\}$, gives rise to the energy diagram shown in Fig.~\ref{fig:energy-diagram} for $B = \SI{5}{\tesla}$. The solid black arrow indicates the first order or direct transition from one spin state to the other. The blurry arrows represent a cascade-like mechanism, where the spin-flip transition is realized in a step-by-step process with smallest possible change in $\Delta M_\mathrm{S}$.

In a many-body theory of the electron-phonon coupling, we need an approximation for the matrix elements between multiplet states, $G_{i,f}^\eta = \langle \Psi_i | H_\mathrm{el.-ph.}^\eta | \Psi_f \rangle$. We calculated these matrix elements by employing the \textit{ab-initio} matrix elements for one-electron states $\ket{\tilde{\psi}_i}$ and contracting the creation and annihilation operators of $H_\mathrm{el.-ph.}$ in Eq.~\eqref{eq:el-ph} and the second quantization representation of the multiplet wave functions $\ket{\Psi_i}$.
To obtain the multiplets represented in second quantization we have used the expression of the basis states in second quantization given in Ref.~\footnotemark[1].
Finally, the fundamental Hamiltonian describing the spin-phonon coupling can be defined as,
\begin{align} \label{eq:Hamiltonian}
H
=
\sum_i E_i C^\dagger_i C_i
+
\sum_{\eta}
\omega_\eta b^\dagger_\eta b_\eta
+
\sum_{i,f,\eta}
G_{if}^\eta \, C^\dagger_f C_i \, (b^\dagger_\eta + b_\eta)
.\end{align}
The first term describes the Fe multiplet states, with energies $E_i$ given by Eq.~\eqref{eq:Fe-Stevens-Hamiltonian-Baumann}. The second term describes the unperturbed phonon system and $\omega_\eta$ are the frequencies of the lattice vibrations obtained by the DFT calculations. Finally, the third term describes the coupling between electrons and phonons. Even if simple, Eq.~\eqref{eq:Hamiltonian} comprises the information about the multiplet nature of the electron states in the Fe adatom and the \textit{ab-initio} details of the substrate electron and phonon structure.

The time evolution of the adatom coupled with the phonon bath, represented by Eq.~\eqref{eq:Hamiltonian}, can be described using a master equation for the density matrix of the adatom multiplet states. For low temperatures considered in the experimental setup~\cite{Paul2017} ($k_\mathrm{B} T \approx \SI{0.1}{\milli\electronvolt}$), it is a good approximation to neglect the scattering to the higher energy states $|\Psi_2\rangle$, $|\Psi_3\rangle$ and $|\Psi_4\rangle$, which we verified by solving the complete master equation. In this case, the spin dynamics are essentially determined by the direct transition between the low energy states $|\Psi_0\rangle$ and $|\Psi_1\rangle$, as indicated by the black arrow in \figurename~\ref{fig:energy-diagram}. This leads to Fermi's Golden Rule for the direct transition~\footnotemark[1],
\begin{align} \label{eq:fermis-golden-rule}
\Gamma_{1\rightarrow0}
\!=\!
2 \pi\!
\sum_{\eta}\!
|{G}_{1,0}^\eta|^2
\left[2n_\mathrm{BE}(\omega_\eta)+1 \right]
\delta(\!E_{1}\!-\!E_{0}\!-\!\omega_\eta)
,\end{align}
with the phonon thermal population given by $n_\mathrm{BE}(\omega_\eta)$.

\begin{figure*}[bth!]
\centering
\captionsetup[subfigure]{labelformat=empty}
\subfloat[\label{fig:lifetime}]{}
\subfloat[\label{fig:TvsF0}]{}
\subfloat[\label{fig:TvsB}]{}
\includegraphics[width=\linewidth]{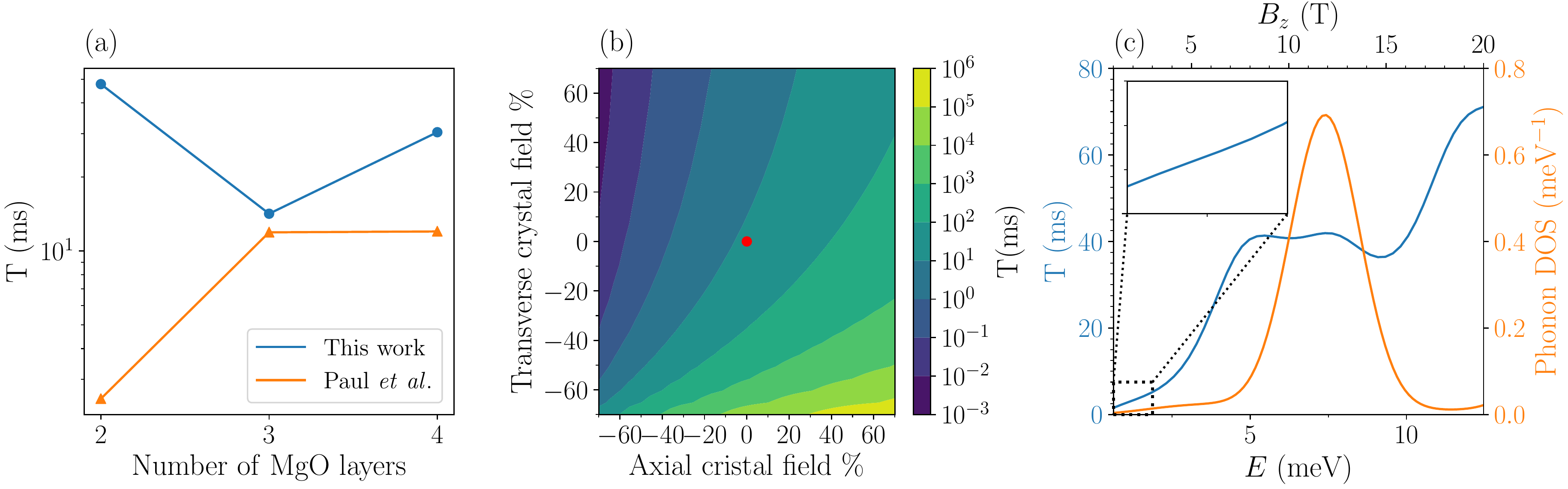}
\caption{(a) Spin lifetime as a function of the MgO coverage. Calculated values (blue) and experimental measurements (orange) from Ref.~\cite{Paul2017} are given by dots, while lines are shown as a guide for the eye. (b) Calculated lifetime as a function of the axial crystal field and transverse crystal field for 3 MLs of MgO. The red dot indicates the crystal field values given in Ref.~\cite{Baumann2015a}. (c) Lifetime as a function of the Zeeman splitting $E$ (left axis) and phonon DOS projected onto the Fe adatom (right axis) for 3 MgO layers.}
\label{fig:results}
\end{figure*}

\figurename~\ref{fig:lifetime} shows the computed spin-flip lifetime using the above equation for MgO coverages ranging from 2 ML to 4 ML at $T = \SI{0}{\kelvin}$ and $B_z = \SI{5}{\tesla}$. The calculated lifetime ranges between $10$ and \SI{50}{\milli\second}. In the case of 2 MLs, the measured spin lifetime is believed to be dominated by the electron-hole pair creation in the substrate~\cite{Paul2017}, which is not included in our model; it is therefore reasonable that our calculated lifetime is nearly an order of magnitude larger than the experimentally measured one. For $>2$ ML coverages, the calculated lifetimes are of the same order as the experimentally measured values~\cite{Paul2017}. Noteworthily, our calculation shows that the lifetime due to the spin-phonon coupling does not change drastically for different coverages, which is also in agreement with the behavior observed experimentally.

On close inspection, we have found that the contributions to the multiplet wave functions $\ket{\Psi_0}$ and $\ket{\Psi_1}$ from states $\ket{M_\mathrm{S}, M_\mathrm{L}} = \ket{\pm2, \pm2}$ are determinant. This reveals that the spin relaxation mechanism is dominated by the overlap of components with $\Delta M_\mathrm{L} = \pm 4$, in particular the one-electron $d_{xy}$ and $d_{x^2-y^2}$ orbitals shown in \figurename~\ref{fig:pdos}. This has important consequences and means that an effective Hamiltonian of the form
\begin{align}
H_\mathrm{eff}
\propto
L_+^4 + L_-^4
\end{align}
captures the main characteristics of the spin-phonon coupling of the Fe adatom.

Not less important is the observation that if TCF is absent ($B_4^4=0$ on Eq.~\eqref{eq:Fe-Stevens-Hamiltonian-Baumann}), the components with $\Delta M_L = \pm 4$ would not be mixed on the multiplet wave functions, preventing the decay of state $|\Psi_1\rangle$  into $|\Psi_0\rangle$. To investigate this aspect, we have computed the spin-lifetime varying both the ACF and the TCF to determine which term influences most the spin-phonon transition. \figurename~\ref{fig:TvsF0} shows that the lifetime is influenced in orders of magnitude by variations not only of the TCF, but also of the ACF. Additionally, \figurename~\ref{fig:TvsF0} reveals that in order to achieve longer spin lifetimes the ACF should be increased, whereas the TCF should be reduced. It is important to point out that the TCF is much smaller than the ACF (see parameters of Eq.\eqref{eq:Fe-Stevens-Hamiltonian-Baumann}) and that \figurename~\ref{fig:TvsF0} shows relative variations of the crystal field parameters. However, small changes on the TCF have a much larger impact on the low-energy spectrum of the adatom shown in \figurename~\ref{fig:energy-diagram}, while variations in the ACF barely influence the low-energy spectrum. Moreover, as suggested in Ref.~\cite{Baumann2015b}, the TCF can be easily modified in experiment, for example applying an oscillating bias voltage, which could have a big impact on the spin-lifetime measurements according to our results.

Another interesting property that is accessible in our model is the influence of the magnetic field on the spin-lifetime. \figurename~\ref{fig:TvsB} shows the calculated lifetime (left axis) as a function of the magnetic field or Zeeman splitting energy. We observe that for magnetic fields lower than \SI{2}{\tesla}, the lifetime increases linearly with the magnetic field, in excellent agreement with previous experimental measurements~\cite{VanWeerdenburg2021}. This originates from the decreasing admixture of the states enabling $\Delta M_L = \pm 4$ transitions in the multiplet wave functions as the magnetic field increases. For this low energy range, the main scattering channel are phonons that are delocalized throughout the entire system.
Our calculations indicate that 
the displacement of substrate Ag atoms do not contribute significantly to the scattering, while displacements of MgO and Fe contribute roughly equally.

Interestingly, the lifetime saturates when the Zeeman splitting increases to match the energy of the localized in-plane mode of the iron adatom (shown by the projected phonon DOS in \figurename~\ref{fig:TvsB}), due to the sharp increase in the density of available final states for that energy range. As such localized vibrations are a common feature of magnetic adatoms, we propose that the experimental detection of such a plateau in the lifetime of the spin transitions is a clear way to prove that the spin-phonon coupling is indeed the primary relaxation mechanism.

In summary, we presented a microscopic theory of the electron-phonon-induced spin-flip transitions of adatoms on surfaces, and we applied it to the Fe/MgO/Ag(100) system. We tackled this challenging problem which typically involves hundreds of atoms by developing a model that combines the multiplet structure of the adatom with the \textit{ab-initio} information about all the electrons and phonons in the system and their coupling. Our analysis shows a good qualitative and order-of-magnitude agreement with available experiments by Paul et al.~\cite{Paul2017}, which demonstrates that the essential features of the problem are successfully captured. In particular, we find that a correct description of the substrate is essential and that an effective local description of the adatom is insufficient.
Moreover, our model allows us to identify the most important components of the multiplet states in terms of their interaction with vibrations, which can help engineer systems with longer spin-relaxation times. Finally, we propose that a saturation of the spin lifetime as a function of the applied magnetic field is a clear fingerprint of a dominant spin-phonon contribution to the relaxation process, and provides a valuable insight for analysing and understanding experimental data.

The authors acknowledge the Department of Education, Universities and Research of the Eusko Jaurlaritza and the University of the Basque Country UPV/EHU (Grant No.\ IT1260-19 and No.\ IT1527-22), the Spanish Ministry of Economy and Competitiveness MINECO (Grants No. FIS2016-75862-P and No. PID2019-103910GB-I00) for financial support. The work of M.d.S.D.\ made use of computational support by CoSeC, the Computational Science Centre for Research Communities, through CCP9. This project has also received funding from the  European Union's Horizon 2020 research and innovation program under the European Research Council (ERC) grant agreement 946629. H.G.-M.\ acknowledges the Spanish Ministry of Economy and Competitiveness MINECO (Grant No. BES-2017-080039) and the Donostia International Physics Center (DIPC) for financial support. Computer facilities were provided by the DIPC and Centro de F{\'i}sica de Materiales.

\bibliographystyle{apsrev4-2}
\bibliography{./bibliografia}

\begin{thebibliography}{6}%
\makeatletter
\providecommand \@ifxundefined [1]{%
 \@ifx{#1\undefined}
}%
\providecommand \@ifnum [1]{%
 \ifnum #1\expandafter \@firstoftwo
 \else \expandafter \@secondoftwo
 \fi
}%
\providecommand \@ifx [1]{%
 \ifx #1\expandafter \@firstoftwo
 \else \expandafter \@secondoftwo
 \fi
}%
\providecommand \natexlab [1]{#1}%
\providecommand \enquote  [1]{``#1''}%
\providecommand \bibnamefont  [1]{#1}%
\providecommand \bibfnamefont [1]{#1}%
\providecommand \citenamefont [1]{#1}%
\providecommand \href@noop [0]{\@secondoftwo}%
\providecommand \href [0]{\begingroup \@sanitize@url \@href}%
\providecommand \@href[1]{\@@startlink{#1}\@@href}%
\providecommand \@@href[1]{\endgroup#1\@@endlink}%
\providecommand \@sanitize@url [0]{\catcode `\\12\catcode `\$12\catcode
  `\&12\catcode `\#12\catcode `\^12\catcode `\_12\catcode `\%12\relax}%
\providecommand \@@startlink[1]{}%
\providecommand \@@endlink[0]{}%
\providecommand \url  [0]{\begingroup\@sanitize@url \@url }%
\providecommand \@url [1]{\endgroup\@href {#1}{\urlprefix }}%
\providecommand \urlprefix  [0]{URL }%
\providecommand \Eprint [0]{\href }%
\providecommand \doibase [0]{https://doi.org/}%
\providecommand \selectlanguage [0]{\@gobble}%
\providecommand \bibinfo  [0]{\@secondoftwo}%
\providecommand \bibfield  [0]{\@secondoftwo}%
\providecommand \translation [1]{[#1]}%
\providecommand \BibitemOpen [0]{}%
\providecommand \bibitemStop [0]{}%
\providecommand \bibitemNoStop [0]{.\EOS\space}%
\providecommand \EOS [0]{\spacefactor3000\relax}%
\providecommand \BibitemShut  [1]{\csname bibitem#1\endcsname}%
\let\auto@bib@innerbib\@empty
\bibitem [{\citenamefont {Garai-Marin}\ \emph {et~al.}(2021)\citenamefont
  {Garai-Marin}, \citenamefont {Iba{\~{n}}ez-Azpiroz}, \citenamefont
  {Garcia-Goiricelaya}, \citenamefont {Gurtubay},\ and\ \citenamefont
  {Eiguren}}]{Garai-Marin2021}%
  \BibitemOpen
  \bibfield  {author} {\bibinfo {author} {\bibfnamefont {H.}~\bibnamefont
  {Garai-Marin}}, \bibinfo {author} {\bibfnamefont {J.}~\bibnamefont
  {Iba{\~{n}}ez-Azpiroz}}, \bibinfo {author} {\bibfnamefont {P.}~\bibnamefont
  {Garcia-Goiricelaya}}, \bibinfo {author} {\bibfnamefont {I.~G.}\ \bibnamefont
  {Gurtubay}},\ and\ \bibinfo {author} {\bibfnamefont {A.}~\bibnamefont
  {Eiguren}},\ }\href {https://doi.org/10.1103/PhysRevB.104.195422} {\bibfield
  {journal} {\bibinfo  {journal} {Physical Review B}\ }\textbf {\bibinfo
  {volume} {104}},\ \bibinfo {pages} {195422} (\bibinfo {year}
  {2021})}\BibitemShut {NoStop}%
\bibitem [{\citenamefont {Soler}\ \emph {et~al.}(2002)\citenamefont {Soler},
  \citenamefont {Artacho}, \citenamefont {Gale}, \citenamefont {Garc{\'{i}}a},
  \citenamefont {Junquera}, \citenamefont {Ordej{\'{o}}n},\ and\ \citenamefont
  {S{\'{a}}nchez-Portal}}]{Soler2002}%
  \BibitemOpen
  \bibfield  {author} {\bibinfo {author} {\bibfnamefont {J.~M.}\ \bibnamefont
  {Soler}}, \bibinfo {author} {\bibfnamefont {E.}~\bibnamefont {Artacho}},
  \bibinfo {author} {\bibfnamefont {J.~D.}\ \bibnamefont {Gale}}, \bibinfo
  {author} {\bibfnamefont {A.}~\bibnamefont {Garc{\'{i}}a}}, \bibinfo {author}
  {\bibfnamefont {J.}~\bibnamefont {Junquera}}, \bibinfo {author}
  {\bibfnamefont {P.}~\bibnamefont {Ordej{\'{o}}n}},\ and\ \bibinfo {author}
  {\bibfnamefont {D.}~\bibnamefont {S{\'{a}}nchez-Portal}},\ }\href
  {https://doi.org/10.1088/0953-8984/14/11/302} {\bibfield  {journal} {\bibinfo
   {journal} {Journal of Physics: Condensed Matter}\ }\textbf {\bibinfo
  {volume} {14}},\ \bibinfo {pages} {2745} (\bibinfo {year}
  {2002})}\BibitemShut {NoStop}%
\bibitem [{\citenamefont {Cuadrado}\ and\ \citenamefont {{I.
  Cerd{\'{a}}}}(2012)}]{Cuadrado2012}%
  \BibitemOpen
  \bibfield  {author} {\bibinfo {author} {\bibfnamefont {R.}~\bibnamefont
  {Cuadrado}}\ and\ \bibinfo {author} {\bibfnamefont {J.}~\bibnamefont {{I.
  Cerd{\'{a}}}}},\ }\href {https://doi.org/10.1088/0953-8984/24/8/086005}
  {\bibfield  {journal} {\bibinfo  {journal} {Journal of Physics: Condensed
  Matter}\ }\textbf {\bibinfo {volume} {24}},\ \bibinfo {pages} {086005}
  (\bibinfo {year} {2012})}\BibitemShut {NoStop}%
\bibitem [{\citenamefont {Perdew}\ \emph {et~al.}(1996)\citenamefont {Perdew},
  \citenamefont {Burke},\ and\ \citenamefont {Ernzerhof}}]{Perdew1996}%
  \BibitemOpen
  \bibfield  {author} {\bibinfo {author} {\bibfnamefont {J.~P.}\ \bibnamefont
  {Perdew}}, \bibinfo {author} {\bibfnamefont {K.}~\bibnamefont {Burke}},\ and\
  \bibinfo {author} {\bibfnamefont {M.}~\bibnamefont {Ernzerhof}},\ }\href
  {https://doi.org/10.1103/PhysRevLett.77.3865} {\bibfield  {journal} {\bibinfo
   {journal} {Physical Review Letters}\ }\textbf {\bibinfo {volume} {77}},\
  \bibinfo {pages} {3865} (\bibinfo {year} {1996})}\BibitemShut {NoStop}%
\bibitem [{\citenamefont {Kleinman}\ and\ \citenamefont
  {Bylander}(1982)}]{Kleinman1982}%
  \BibitemOpen
  \bibfield  {author} {\bibinfo {author} {\bibfnamefont {L.}~\bibnamefont
  {Kleinman}}\ and\ \bibinfo {author} {\bibfnamefont {D.~M.}\ \bibnamefont
  {Bylander}},\ }\href {https://doi.org/10.1103/PhysRevLett.48.1425} {\bibfield
   {journal} {\bibinfo  {journal} {Physical Review Letters}\ }\textbf {\bibinfo
  {volume} {48}},\ \bibinfo {pages} {1425} (\bibinfo {year}
  {1982})}\BibitemShut {NoStop}%
\bibitem [{\citenamefont {Breuer}\ and\ \citenamefont
  {Petruccione}(2007)}]{Breuer2007}%
  \BibitemOpen
  \bibfield  {author} {\bibinfo {author} {\bibfnamefont {H.-P.}\ \bibnamefont
  {Breuer}}\ and\ \bibinfo {author} {\bibfnamefont {F.}~\bibnamefont
  {Petruccione}},\ }\href
  {https://doi.org/10.1093/acprof:oso/9780199213900.001.0001} {\emph {\bibinfo
  {title} {{The Theory of Open Quantum Systems}}}}\ (\bibinfo  {publisher}
  {Oxford University Press},\ \bibinfo {year} {2007})\BibitemShut {NoStop}%
\end{thebibliography}%


\begin{thebibliography}{57}%
\makeatletter
\providecommand \@ifxundefined [1]{%
 \@ifx{#1\undefined}
}%
\providecommand \@ifnum [1]{%
 \ifnum #1\expandafter \@firstoftwo
 \else \expandafter \@secondoftwo
 \fi
}%
\providecommand \@ifx [1]{%
 \ifx #1\expandafter \@firstoftwo
 \else \expandafter \@secondoftwo
 \fi
}%
\providecommand \natexlab [1]{#1}%
\providecommand \enquote  [1]{``#1''}%
\providecommand \bibnamefont  [1]{#1}%
\providecommand \bibfnamefont [1]{#1}%
\providecommand \citenamefont [1]{#1}%
\providecommand \href@noop [0]{\@secondoftwo}%
\providecommand \href [0]{\begingroup \@sanitize@url \@href}%
\providecommand \@href[1]{\@@startlink{#1}\@@href}%
\providecommand \@@href[1]{\endgroup#1\@@endlink}%
\providecommand \@sanitize@url [0]{\catcode `\\12\catcode `\$12\catcode
  `\&12\catcode `\#12\catcode `\^12\catcode `\_12\catcode `\%12\relax}%
\providecommand \@@startlink[1]{}%
\providecommand \@@endlink[0]{}%
\providecommand \url  [0]{\begingroup\@sanitize@url \@url }%
\providecommand \@url [1]{\endgroup\@href {#1}{\urlprefix }}%
\providecommand \urlprefix  [0]{URL }%
\providecommand \Eprint [0]{\href }%
\providecommand \doibase [0]{https://doi.org/}%
\providecommand \selectlanguage [0]{\@gobble}%
\providecommand \bibinfo  [0]{\@secondoftwo}%
\providecommand \bibfield  [0]{\@secondoftwo}%
\providecommand \translation [1]{[#1]}%
\providecommand \BibitemOpen [0]{}%
\providecommand \bibitemStop [0]{}%
\providecommand \bibitemNoStop [0]{.\EOS\space}%
\providecommand \EOS [0]{\spacefactor3000\relax}%
\providecommand \BibitemShut  [1]{\csname bibitem#1\endcsname}%
\let\auto@bib@innerbib\@empty
\bibitem [{\citenamefont {Heinrich}\ \emph {et~al.}(2004)\citenamefont
  {Heinrich}, \citenamefont {Gupta}, \citenamefont {Lutz},\ and\ \citenamefont
  {Eigler}}]{Heinrich2004}%
  \BibitemOpen
  \bibfield  {author} {\bibinfo {author} {\bibfnamefont {A.~J.}\ \bibnamefont
  {Heinrich}}, \bibinfo {author} {\bibfnamefont {J.~A.}\ \bibnamefont {Gupta}},
  \bibinfo {author} {\bibfnamefont {C.~P.}\ \bibnamefont {Lutz}},\ and\
  \bibinfo {author} {\bibfnamefont {D.~M.}\ \bibnamefont {Eigler}},\ }\href
  {https://doi.org/10.1126/science.1101077} {\bibfield  {journal} {\bibinfo
  {journal} {Science}\ }\textbf {\bibinfo {volume} {306}},\ \bibinfo {pages}
  {466} (\bibinfo {year} {2004})}\BibitemShut {NoStop}%
\bibitem [{\citenamefont {Loth}\ \emph
  {et~al.}(2010{\natexlab{a}})\citenamefont {Loth}, \citenamefont {Lutz},\ and\
  \citenamefont {Heinrich}}]{Loth2010a}%
  \BibitemOpen
  \bibfield  {author} {\bibinfo {author} {\bibfnamefont {S.}~\bibnamefont
  {Loth}}, \bibinfo {author} {\bibfnamefont {C.~P.}\ \bibnamefont {Lutz}},\
  and\ \bibinfo {author} {\bibfnamefont {A.~J.}\ \bibnamefont {Heinrich}},\
  }\href {https://doi.org/10.1088/1367-2630/12/12/125021} {\bibfield  {journal}
  {\bibinfo  {journal} {New Journal of Physics}\ }\textbf {\bibinfo {volume}
  {12}},\ \bibinfo {pages} {125021} (\bibinfo {year}
  {2010}{\natexlab{a}})}\BibitemShut {NoStop}%
\bibitem [{\citenamefont {Ternes}(2015)}]{Ternes2015}%
  \BibitemOpen
  \bibfield  {author} {\bibinfo {author} {\bibfnamefont {M.}~\bibnamefont
  {Ternes}},\ }\href {https://doi.org/10.1088/1367-2630/17/6/063016} {\bibfield
   {journal} {\bibinfo  {journal} {New Journal of Physics}\ }\textbf {\bibinfo
  {volume} {17}},\ \bibinfo {pages} {063016} (\bibinfo {year}
  {2015})}\BibitemShut {NoStop}%
\bibitem [{\citenamefont {Otte}\ \emph {et~al.}(2008)\citenamefont {Otte},
  \citenamefont {Ternes}, \citenamefont {von Bergmann}, \citenamefont {Loth},
  \citenamefont {Brune}, \citenamefont {Lutz}, \citenamefont {Hirjibehedin},\
  and\ \citenamefont {Heinrich}}]{Otte2008}%
  \BibitemOpen
  \bibfield  {author} {\bibinfo {author} {\bibfnamefont {A.~F.}\ \bibnamefont
  {Otte}}, \bibinfo {author} {\bibfnamefont {M.}~\bibnamefont {Ternes}},
  \bibinfo {author} {\bibfnamefont {K.}~\bibnamefont {von Bergmann}}, \bibinfo
  {author} {\bibfnamefont {S.}~\bibnamefont {Loth}}, \bibinfo {author}
  {\bibfnamefont {H.}~\bibnamefont {Brune}}, \bibinfo {author} {\bibfnamefont
  {C.~P.}\ \bibnamefont {Lutz}}, \bibinfo {author} {\bibfnamefont {C.~F.}\
  \bibnamefont {Hirjibehedin}},\ and\ \bibinfo {author} {\bibfnamefont {A.~J.}\
  \bibnamefont {Heinrich}},\ }\href {https://doi.org/10.1038/nphys1072}
  {\bibfield  {journal} {\bibinfo  {journal} {Nature Physics}\ }\textbf
  {\bibinfo {volume} {4}},\ \bibinfo {pages} {847} (\bibinfo {year}
  {2008})}\BibitemShut {NoStop}%
\bibitem [{\citenamefont {Meier}\ \emph {et~al.}(2008)\citenamefont {Meier},
  \citenamefont {Zhou}, \citenamefont {Wiebe},\ and\ \citenamefont
  {Wiesendanger}}]{Meier2008}%
  \BibitemOpen
  \bibfield  {author} {\bibinfo {author} {\bibfnamefont {F.}~\bibnamefont
  {Meier}}, \bibinfo {author} {\bibfnamefont {L.}~\bibnamefont {Zhou}},
  \bibinfo {author} {\bibfnamefont {J.}~\bibnamefont {Wiebe}},\ and\ \bibinfo
  {author} {\bibfnamefont {R.}~\bibnamefont {Wiesendanger}},\ }\href
  {https://doi.org/10.1126/science.1154415} {\bibfield  {journal} {\bibinfo
  {journal} {Science}\ }\textbf {\bibinfo {volume} {320}},\ \bibinfo {pages}
  {82} (\bibinfo {year} {2008})}\BibitemShut {NoStop}%
\bibitem [{\citenamefont {Bouaziz}\ \emph {et~al.}(2020)\citenamefont
  {Bouaziz}, \citenamefont {Iba{\~{n}}ez-Azpiroz}, \citenamefont
  {Guimar{\~{a}}es},\ and\ \citenamefont {Lounis}}]{Bouaziz2020}%
  \BibitemOpen
  \bibfield  {author} {\bibinfo {author} {\bibfnamefont {J.}~\bibnamefont
  {Bouaziz}}, \bibinfo {author} {\bibfnamefont {J.}~\bibnamefont
  {Iba{\~{n}}ez-Azpiroz}}, \bibinfo {author} {\bibfnamefont {F.~S.~M.}\
  \bibnamefont {Guimar{\~{a}}es}},\ and\ \bibinfo {author} {\bibfnamefont
  {S.}~\bibnamefont {Lounis}},\ }\href
  {https://doi.org/10.1103/PhysRevResearch.2.043357} {\bibfield  {journal}
  {\bibinfo  {journal} {Physical Review Research}\ }\textbf {\bibinfo {volume}
  {2}},\ \bibinfo {pages} {043357} (\bibinfo {year} {2020})}\BibitemShut
  {NoStop}%
\bibitem [{\citenamefont {Delgado}\ and\ \citenamefont
  {Fern{\'{a}}ndez-Rossier}(2017)}]{Delgado2017}%
  \BibitemOpen
  \bibfield  {author} {\bibinfo {author} {\bibfnamefont {F.}~\bibnamefont
  {Delgado}}\ and\ \bibinfo {author} {\bibfnamefont {J.}~\bibnamefont
  {Fern{\'{a}}ndez-Rossier}},\ }\href
  {https://doi.org/10.1016/j.progsurf.2016.12.001} {\bibfield  {journal}
  {\bibinfo  {journal} {Progress in Surface Science}\ }\textbf {\bibinfo
  {volume} {92}},\ \bibinfo {pages} {40} (\bibinfo {year} {2017})}\BibitemShut
  {NoStop}%
\bibitem [{\citenamefont {Donati}\ \emph {et~al.}(2016)\citenamefont {Donati},
  \citenamefont {Rusponi}, \citenamefont {Stepanow}, \citenamefont {Wackerlin},
  \citenamefont {Singha}, \citenamefont {Persichetti}, \citenamefont {Baltic},
  \citenamefont {Diller}, \citenamefont {Patthey}, \citenamefont {Fernandes},
  \citenamefont {Dreiser}, \citenamefont {{\v{S}}ljivan{\v{c}}anin},
  \citenamefont {Kummer}, \citenamefont {Nistor}, \citenamefont {Gambardella},\
  and\ \citenamefont {Brune}}]{Donati2016}%
  \BibitemOpen
  \bibfield  {author} {\bibinfo {author} {\bibfnamefont {F.}~\bibnamefont
  {Donati}}, \bibinfo {author} {\bibfnamefont {S.}~\bibnamefont {Rusponi}},
  \bibinfo {author} {\bibfnamefont {S.}~\bibnamefont {Stepanow}}, \bibinfo
  {author} {\bibfnamefont {C.}~\bibnamefont {Wackerlin}}, \bibinfo {author}
  {\bibfnamefont {A.}~\bibnamefont {Singha}}, \bibinfo {author} {\bibfnamefont
  {L.}~\bibnamefont {Persichetti}}, \bibinfo {author} {\bibfnamefont
  {R.}~\bibnamefont {Baltic}}, \bibinfo {author} {\bibfnamefont
  {K.}~\bibnamefont {Diller}}, \bibinfo {author} {\bibfnamefont
  {F.}~\bibnamefont {Patthey}}, \bibinfo {author} {\bibfnamefont
  {E.}~\bibnamefont {Fernandes}}, \bibinfo {author} {\bibfnamefont
  {J.}~\bibnamefont {Dreiser}}, \bibinfo {author} {\bibfnamefont
  {{\v{Z}}.}~\bibnamefont {{\v{S}}ljivan{\v{c}}anin}}, \bibinfo {author}
  {\bibfnamefont {K.}~\bibnamefont {Kummer}}, \bibinfo {author} {\bibfnamefont
  {C.}~\bibnamefont {Nistor}}, \bibinfo {author} {\bibfnamefont
  {P.}~\bibnamefont {Gambardella}},\ and\ \bibinfo {author} {\bibfnamefont
  {H.}~\bibnamefont {Brune}},\ }\href {https://doi.org/10.1126/science.aad9898}
  {\bibfield  {journal} {\bibinfo  {journal} {Science}\ }\textbf {\bibinfo
  {volume} {352}},\ \bibinfo {pages} {318} (\bibinfo {year}
  {2016})}\BibitemShut {NoStop}%
\bibitem [{\citenamefont {Paul}\ \emph {et~al.}(2017)\citenamefont {Paul},
  \citenamefont {Yang}, \citenamefont {Baumann}, \citenamefont {Romming},
  \citenamefont {Choi}, \citenamefont {Lutz},\ and\ \citenamefont
  {Heinrich}}]{Paul2017}%
  \BibitemOpen
  \bibfield  {author} {\bibinfo {author} {\bibfnamefont {W.}~\bibnamefont
  {Paul}}, \bibinfo {author} {\bibfnamefont {K.}~\bibnamefont {Yang}}, \bibinfo
  {author} {\bibfnamefont {S.}~\bibnamefont {Baumann}}, \bibinfo {author}
  {\bibfnamefont {N.}~\bibnamefont {Romming}}, \bibinfo {author} {\bibfnamefont
  {T.}~\bibnamefont {Choi}}, \bibinfo {author} {\bibfnamefont {C.~P.}\
  \bibnamefont {Lutz}},\ and\ \bibinfo {author} {\bibfnamefont {A.~J.}\
  \bibnamefont {Heinrich}},\ }\href {https://doi.org/10.1038/nphys3965}
  {\bibfield  {journal} {\bibinfo  {journal} {Nature Physics}\ }\textbf
  {\bibinfo {volume} {13}},\ \bibinfo {pages} {403} (\bibinfo {year}
  {2017})}\BibitemShut {NoStop}%
\bibitem [{\citenamefont {Natterer}\ \emph {et~al.}(2018)\citenamefont
  {Natterer}, \citenamefont {Donati}, \citenamefont {Patthey},\ and\
  \citenamefont {Brune}}]{Natterer2018}%
  \BibitemOpen
  \bibfield  {author} {\bibinfo {author} {\bibfnamefont {F.~D.}\ \bibnamefont
  {Natterer}}, \bibinfo {author} {\bibfnamefont {F.}~\bibnamefont {Donati}},
  \bibinfo {author} {\bibfnamefont {F.}~\bibnamefont {Patthey}},\ and\ \bibinfo
  {author} {\bibfnamefont {H.}~\bibnamefont {Brune}},\ }\href
  {https://doi.org/10.1103/PhysRevLett.121.027201} {\bibfield  {journal}
  {\bibinfo  {journal} {Physical Review Letters}\ }\textbf {\bibinfo {volume}
  {121}},\ \bibinfo {pages} {027201} (\bibinfo {year} {2018})}\BibitemShut
  {NoStop}%
\bibitem [{\citenamefont {Miyamachi}\ \emph {et~al.}(2013)\citenamefont
  {Miyamachi}, \citenamefont {Schuh}, \citenamefont {M{\"{a}}rkl},
  \citenamefont {Bresch}, \citenamefont {Balashov}, \citenamefont
  {St{\"{o}}hr}, \citenamefont {Karlewski}, \citenamefont {Andr{\'{e}}},
  \citenamefont {Marthaler}, \citenamefont {Hoffmann}, \citenamefont
  {Geilhufe}, \citenamefont {Ostanin}, \citenamefont {Hergert}, \citenamefont
  {Mertig}, \citenamefont {Sch{\"{o}}n}, \citenamefont {Ernst},\ and\
  \citenamefont {Wulfhekel}}]{Miyamachi2013}%
  \BibitemOpen
  \bibfield  {author} {\bibinfo {author} {\bibfnamefont {T.}~\bibnamefont
  {Miyamachi}}, \bibinfo {author} {\bibfnamefont {T.}~\bibnamefont {Schuh}},
  \bibinfo {author} {\bibfnamefont {T.}~\bibnamefont {M{\"{a}}rkl}}, \bibinfo
  {author} {\bibfnamefont {C.}~\bibnamefont {Bresch}}, \bibinfo {author}
  {\bibfnamefont {T.}~\bibnamefont {Balashov}}, \bibinfo {author}
  {\bibfnamefont {A.}~\bibnamefont {St{\"{o}}hr}}, \bibinfo {author}
  {\bibfnamefont {C.}~\bibnamefont {Karlewski}}, \bibinfo {author}
  {\bibfnamefont {S.}~\bibnamefont {Andr{\'{e}}}}, \bibinfo {author}
  {\bibfnamefont {M.}~\bibnamefont {Marthaler}}, \bibinfo {author}
  {\bibfnamefont {M.}~\bibnamefont {Hoffmann}}, \bibinfo {author}
  {\bibfnamefont {M.}~\bibnamefont {Geilhufe}}, \bibinfo {author}
  {\bibfnamefont {S.}~\bibnamefont {Ostanin}}, \bibinfo {author} {\bibfnamefont
  {W.}~\bibnamefont {Hergert}}, \bibinfo {author} {\bibfnamefont
  {I.}~\bibnamefont {Mertig}}, \bibinfo {author} {\bibfnamefont
  {G.}~\bibnamefont {Sch{\"{o}}n}}, \bibinfo {author} {\bibfnamefont
  {A.}~\bibnamefont {Ernst}},\ and\ \bibinfo {author} {\bibfnamefont
  {W.}~\bibnamefont {Wulfhekel}},\ }\href {https://doi.org/10.1038/nature12759}
  {\bibfield  {journal} {\bibinfo  {journal} {Nature}\ }\textbf {\bibinfo
  {volume} {503}},\ \bibinfo {pages} {242} (\bibinfo {year}
  {2013})}\BibitemShut {NoStop}%
\bibitem [{\citenamefont {Rau}\ \emph {et~al.}(2014)\citenamefont {Rau},
  \citenamefont {Baumann}, \citenamefont {Rusponi}, \citenamefont {Donati},
  \citenamefont {Stepanow}, \citenamefont {Gragnaniello}, \citenamefont
  {Dreiser}, \citenamefont {Piamonteze}, \citenamefont {Nolting}, \citenamefont
  {Gangopadhyay}, \citenamefont {Albertini}, \citenamefont {Macfarlane},
  \citenamefont {Lutz}, \citenamefont {Jones}, \citenamefont {Gambardella},
  \citenamefont {Heinrich},\ and\ \citenamefont {Brune}}]{Rau2014}%
  \BibitemOpen
  \bibfield  {author} {\bibinfo {author} {\bibfnamefont {I.~G.}\ \bibnamefont
  {Rau}}, \bibinfo {author} {\bibfnamefont {S.}~\bibnamefont {Baumann}},
  \bibinfo {author} {\bibfnamefont {S.}~\bibnamefont {Rusponi}}, \bibinfo
  {author} {\bibfnamefont {F.}~\bibnamefont {Donati}}, \bibinfo {author}
  {\bibfnamefont {S.}~\bibnamefont {Stepanow}}, \bibinfo {author}
  {\bibfnamefont {L.}~\bibnamefont {Gragnaniello}}, \bibinfo {author}
  {\bibfnamefont {J.}~\bibnamefont {Dreiser}}, \bibinfo {author} {\bibfnamefont
  {C.}~\bibnamefont {Piamonteze}}, \bibinfo {author} {\bibfnamefont
  {F.}~\bibnamefont {Nolting}}, \bibinfo {author} {\bibfnamefont
  {S.}~\bibnamefont {Gangopadhyay}}, \bibinfo {author} {\bibfnamefont {O.~R.}\
  \bibnamefont {Albertini}}, \bibinfo {author} {\bibfnamefont {R.~M.}\
  \bibnamefont {Macfarlane}}, \bibinfo {author} {\bibfnamefont {C.~P.}\
  \bibnamefont {Lutz}}, \bibinfo {author} {\bibfnamefont {B.~A.}\ \bibnamefont
  {Jones}}, \bibinfo {author} {\bibfnamefont {P.}~\bibnamefont {Gambardella}},
  \bibinfo {author} {\bibfnamefont {A.~J.}\ \bibnamefont {Heinrich}},\ and\
  \bibinfo {author} {\bibfnamefont {H.}~\bibnamefont {Brune}},\ }\href
  {https://doi.org/10.1126/science.1252841} {\bibfield  {journal} {\bibinfo
  {journal} {Science}\ }\textbf {\bibinfo {volume} {344}},\ \bibinfo {pages}
  {988} (\bibinfo {year} {2014})}\BibitemShut {NoStop}%
\bibitem [{\citenamefont {Loth}\ \emph
  {et~al.}(2010{\natexlab{b}})\citenamefont {Loth}, \citenamefont {von
  Bergmann}, \citenamefont {Ternes}, \citenamefont {Otte}, \citenamefont
  {Lutz},\ and\ \citenamefont {Heinrich}}]{Loth2010}%
  \BibitemOpen
  \bibfield  {author} {\bibinfo {author} {\bibfnamefont {S.}~\bibnamefont
  {Loth}}, \bibinfo {author} {\bibfnamefont {K.}~\bibnamefont {von Bergmann}},
  \bibinfo {author} {\bibfnamefont {M.}~\bibnamefont {Ternes}}, \bibinfo
  {author} {\bibfnamefont {A.~F.}\ \bibnamefont {Otte}}, \bibinfo {author}
  {\bibfnamefont {C.~P.}\ \bibnamefont {Lutz}},\ and\ \bibinfo {author}
  {\bibfnamefont {A.~J.}\ \bibnamefont {Heinrich}},\ }\href
  {https://doi.org/10.1038/nphys1616} {\bibfield  {journal} {\bibinfo
  {journal} {Nature Physics}\ }\textbf {\bibinfo {volume} {6}},\ \bibinfo
  {pages} {340} (\bibinfo {year} {2010}{\natexlab{b}})}\BibitemShut {NoStop}%
\bibitem [{\citenamefont {Natterer}\ \emph {et~al.}(2017)\citenamefont
  {Natterer}, \citenamefont {Yang}, \citenamefont {Paul}, \citenamefont
  {Willke}, \citenamefont {Choi}, \citenamefont {Greber}, \citenamefont
  {Heinrich},\ and\ \citenamefont {Lutz}}]{Natterer2017}%
  \BibitemOpen
  \bibfield  {author} {\bibinfo {author} {\bibfnamefont {F.~D.}\ \bibnamefont
  {Natterer}}, \bibinfo {author} {\bibfnamefont {K.}~\bibnamefont {Yang}},
  \bibinfo {author} {\bibfnamefont {W.}~\bibnamefont {Paul}}, \bibinfo {author}
  {\bibfnamefont {P.}~\bibnamefont {Willke}}, \bibinfo {author} {\bibfnamefont
  {T.}~\bibnamefont {Choi}}, \bibinfo {author} {\bibfnamefont {T.}~\bibnamefont
  {Greber}}, \bibinfo {author} {\bibfnamefont {A.~J.}\ \bibnamefont
  {Heinrich}},\ and\ \bibinfo {author} {\bibfnamefont {C.~P.}\ \bibnamefont
  {Lutz}},\ }\href {https://doi.org/10.1038/nature21371} {\bibfield  {journal}
  {\bibinfo  {journal} {Nature}\ }\textbf {\bibinfo {volume} {543}},\ \bibinfo
  {pages} {226} (\bibinfo {year} {2017})}\BibitemShut {NoStop}%
\bibitem [{\citenamefont
  {Fern{\'{a}}ndez-Rossier}(2009)}]{Fernandez-Rossier2009}%
  \BibitemOpen
  \bibfield  {author} {\bibinfo {author} {\bibfnamefont {J.}~\bibnamefont
  {Fern{\'{a}}ndez-Rossier}},\ }\href
  {https://doi.org/10.1103/PhysRevLett.102.256802} {\bibfield  {journal}
  {\bibinfo  {journal} {Physical Review Letters}\ }\textbf {\bibinfo {volume}
  {102}},\ \bibinfo {pages} {256802} (\bibinfo {year} {2009})}\BibitemShut
  {NoStop}%
\bibitem [{\citenamefont {Lorente}\ and\ \citenamefont
  {Gauyacq}(2009)}]{Lorente2009}%
  \BibitemOpen
  \bibfield  {author} {\bibinfo {author} {\bibfnamefont {N.}~\bibnamefont
  {Lorente}}\ and\ \bibinfo {author} {\bibfnamefont {J.-P.}\ \bibnamefont
  {Gauyacq}},\ }\href {https://doi.org/10.1103/PhysRevLett.103.176601}
  {\bibfield  {journal} {\bibinfo  {journal} {Physical Review Letters}\
  }\textbf {\bibinfo {volume} {103}},\ \bibinfo {pages} {176601} (\bibinfo
  {year} {2009})}\BibitemShut {NoStop}%
\bibitem [{\citenamefont {Baumann}\ \emph {et~al.}(2015)\citenamefont
  {Baumann}, \citenamefont {Paul}, \citenamefont {Choi}, \citenamefont {Lutz},
  \citenamefont {Ardavan},\ and\ \citenamefont {Heinrich}}]{Baumann2015b}%
  \BibitemOpen
  \bibfield  {author} {\bibinfo {author} {\bibfnamefont {S.}~\bibnamefont
  {Baumann}}, \bibinfo {author} {\bibfnamefont {W.}~\bibnamefont {Paul}},
  \bibinfo {author} {\bibfnamefont {T.}~\bibnamefont {Choi}}, \bibinfo {author}
  {\bibfnamefont {C.~P.}\ \bibnamefont {Lutz}}, \bibinfo {author}
  {\bibfnamefont {A.}~\bibnamefont {Ardavan}},\ and\ \bibinfo {author}
  {\bibfnamefont {A.~J.}\ \bibnamefont {Heinrich}},\ }\href
  {https://doi.org/10.1126/science.aac8703} {\bibfield  {journal} {\bibinfo
  {journal} {Science}\ }\textbf {\bibinfo {volume} {350}},\ \bibinfo {pages}
  {417} (\bibinfo {year} {2015})}\BibitemShut {NoStop}%
\bibitem [{\citenamefont {Willke}\ \emph {et~al.}(2018)\citenamefont {Willke},
  \citenamefont {Bae}, \citenamefont {Yang}, \citenamefont {Lado},
  \citenamefont {Ferr{\'{o}}n}, \citenamefont {Choi}, \citenamefont {Ardavan},
  \citenamefont {Fern{\'{a}}ndez-Rossier}, \citenamefont {Heinrich},\ and\
  \citenamefont {Lutz}}]{Willke2018}%
  \BibitemOpen
  \bibfield  {author} {\bibinfo {author} {\bibfnamefont {P.}~\bibnamefont
  {Willke}}, \bibinfo {author} {\bibfnamefont {Y.}~\bibnamefont {Bae}},
  \bibinfo {author} {\bibfnamefont {K.}~\bibnamefont {Yang}}, \bibinfo {author}
  {\bibfnamefont {J.~L.}\ \bibnamefont {Lado}}, \bibinfo {author}
  {\bibfnamefont {A.}~\bibnamefont {Ferr{\'{o}}n}}, \bibinfo {author}
  {\bibfnamefont {T.}~\bibnamefont {Choi}}, \bibinfo {author} {\bibfnamefont
  {A.}~\bibnamefont {Ardavan}}, \bibinfo {author} {\bibfnamefont
  {J.}~\bibnamefont {Fern{\'{a}}ndez-Rossier}}, \bibinfo {author}
  {\bibfnamefont {A.~J.}\ \bibnamefont {Heinrich}},\ and\ \bibinfo {author}
  {\bibfnamefont {C.~P.}\ \bibnamefont {Lutz}},\ }\href
  {https://doi.org/10.1126/science.aat7047} {\bibfield  {journal} {\bibinfo
  {journal} {Science}\ }\textbf {\bibinfo {volume} {362}},\ \bibinfo {pages}
  {336} (\bibinfo {year} {2018})}\BibitemShut {NoStop}%
\bibitem [{\citenamefont {Yang}\ \emph {et~al.}(2018)\citenamefont {Yang},
  \citenamefont {Willke}, \citenamefont {Bae}, \citenamefont {Ferr{\'{o}}n},
  \citenamefont {Lado}, \citenamefont {Ardavan}, \citenamefont
  {Fern{\'{a}}ndez-Rossier}, \citenamefont {Heinrich},\ and\ \citenamefont
  {Lutz}}]{Yang2018}%
  \BibitemOpen
  \bibfield  {author} {\bibinfo {author} {\bibfnamefont {K.}~\bibnamefont
  {Yang}}, \bibinfo {author} {\bibfnamefont {P.}~\bibnamefont {Willke}},
  \bibinfo {author} {\bibfnamefont {Y.}~\bibnamefont {Bae}}, \bibinfo {author}
  {\bibfnamefont {A.}~\bibnamefont {Ferr{\'{o}}n}}, \bibinfo {author}
  {\bibfnamefont {J.~L.}\ \bibnamefont {Lado}}, \bibinfo {author}
  {\bibfnamefont {A.}~\bibnamefont {Ardavan}}, \bibinfo {author} {\bibfnamefont
  {J.}~\bibnamefont {Fern{\'{a}}ndez-Rossier}}, \bibinfo {author}
  {\bibfnamefont {A.~J.}\ \bibnamefont {Heinrich}},\ and\ \bibinfo {author}
  {\bibfnamefont {C.~P.}\ \bibnamefont {Lutz}},\ }\href
  {https://doi.org/10.1038/s41565-018-0296-7} {\bibfield  {journal} {\bibinfo
  {journal} {Nature Nanotechnology}\ }\textbf {\bibinfo {volume} {13}},\
  \bibinfo {pages} {1120} (\bibinfo {year} {2018})}\BibitemShut {NoStop}%
\bibitem [{\citenamefont {Yang}\ \emph
  {et~al.}(2019{\natexlab{a}})\citenamefont {Yang}, \citenamefont {Paul},
  \citenamefont {Phark}, \citenamefont {Willke}, \citenamefont {Bae},
  \citenamefont {Choi}, \citenamefont {Esat}, \citenamefont {Ardavan},
  \citenamefont {Heinrich},\ and\ \citenamefont {Lutz}}]{Yang2019}%
  \BibitemOpen
  \bibfield  {author} {\bibinfo {author} {\bibfnamefont {K.}~\bibnamefont
  {Yang}}, \bibinfo {author} {\bibfnamefont {W.}~\bibnamefont {Paul}}, \bibinfo
  {author} {\bibfnamefont {S.~H.}\ \bibnamefont {Phark}}, \bibinfo {author}
  {\bibfnamefont {P.}~\bibnamefont {Willke}}, \bibinfo {author} {\bibfnamefont
  {Y.}~\bibnamefont {Bae}}, \bibinfo {author} {\bibfnamefont {T.}~\bibnamefont
  {Choi}}, \bibinfo {author} {\bibfnamefont {T.}~\bibnamefont {Esat}}, \bibinfo
  {author} {\bibfnamefont {A.}~\bibnamefont {Ardavan}}, \bibinfo {author}
  {\bibfnamefont {A.~J.}\ \bibnamefont {Heinrich}},\ and\ \bibinfo {author}
  {\bibfnamefont {C.~P.}\ \bibnamefont {Lutz}},\ }\href
  {https://doi.org/10.1126/science.aay6779} {\bibfield  {journal} {\bibinfo
  {journal} {Science}\ }\textbf {\bibinfo {volume} {366}},\ \bibinfo {pages}
  {509} (\bibinfo {year} {2019}{\natexlab{a}})}\BibitemShut {NoStop}%
\bibitem [{\citenamefont {Yang}\ \emph
  {et~al.}(2019{\natexlab{b}})\citenamefont {Yang}, \citenamefont {Paul},
  \citenamefont {Natterer}, \citenamefont {Lado}, \citenamefont {Bae},
  \citenamefont {Willke}, \citenamefont {Choi}, \citenamefont {Ferr{\'{o}}n},
  \citenamefont {Fern{\'{a}}ndez-Rossier}, \citenamefont {Heinrich},\ and\
  \citenamefont {Lutz}}]{Yang2019a}%
  \BibitemOpen
  \bibfield  {author} {\bibinfo {author} {\bibfnamefont {K.}~\bibnamefont
  {Yang}}, \bibinfo {author} {\bibfnamefont {W.}~\bibnamefont {Paul}}, \bibinfo
  {author} {\bibfnamefont {F.~D.}\ \bibnamefont {Natterer}}, \bibinfo {author}
  {\bibfnamefont {J.~L.}\ \bibnamefont {Lado}}, \bibinfo {author}
  {\bibfnamefont {Y.}~\bibnamefont {Bae}}, \bibinfo {author} {\bibfnamefont
  {P.}~\bibnamefont {Willke}}, \bibinfo {author} {\bibfnamefont
  {T.}~\bibnamefont {Choi}}, \bibinfo {author} {\bibfnamefont {A.}~\bibnamefont
  {Ferr{\'{o}}n}}, \bibinfo {author} {\bibfnamefont {J.}~\bibnamefont
  {Fern{\'{a}}ndez-Rossier}}, \bibinfo {author} {\bibfnamefont {A.~J.}\
  \bibnamefont {Heinrich}},\ and\ \bibinfo {author} {\bibfnamefont {C.~P.}\
  \bibnamefont {Lutz}},\ }\href
  {https://doi.org/10.1103/PHYSREVLETT.122.227203/FIGURES/4/MEDIUM} {\bibfield
  {journal} {\bibinfo  {journal} {Physical Review Letters}\ }\textbf {\bibinfo
  {volume} {122}},\ \bibinfo {pages} {227203} (\bibinfo {year}
  {2019}{\natexlab{b}})}\BibitemShut {NoStop}%
\bibitem [{\citenamefont {{Reina G{\'{a}}lvez}}\ \emph
  {et~al.}(2019)\citenamefont {{Reina G{\'{a}}lvez}}, \citenamefont {Wolf},
  \citenamefont {Delgado},\ and\ \citenamefont {Lorente}}]{ReinaGalvez2019}%
  \BibitemOpen
  \bibfield  {author} {\bibinfo {author} {\bibfnamefont {J.}~\bibnamefont
  {{Reina G{\'{a}}lvez}}}, \bibinfo {author} {\bibfnamefont {C.}~\bibnamefont
  {Wolf}}, \bibinfo {author} {\bibfnamefont {F.}~\bibnamefont {Delgado}},\ and\
  \bibinfo {author} {\bibfnamefont {N.}~\bibnamefont {Lorente}},\ }\href
  {https://doi.org/10.1103/PhysRevB.100.035411} {\bibfield  {journal} {\bibinfo
   {journal} {Physical Review B}\ }\textbf {\bibinfo {volume} {100}},\ \bibinfo
  {pages} {035411} (\bibinfo {year} {2019})}\BibitemShut {NoStop}%
\bibitem [{\citenamefont {Seifert}\ \emph {et~al.}(2020)\citenamefont
  {Seifert}, \citenamefont {Kovarik}, \citenamefont {Juraschek}, \citenamefont
  {Spaldin}, \citenamefont {Gambardella},\ and\ \citenamefont
  {Stepanow}}]{Seifert2020}%
  \BibitemOpen
  \bibfield  {author} {\bibinfo {author} {\bibfnamefont {T.~S.}\ \bibnamefont
  {Seifert}}, \bibinfo {author} {\bibfnamefont {S.}~\bibnamefont {Kovarik}},
  \bibinfo {author} {\bibfnamefont {D.~M.}\ \bibnamefont {Juraschek}}, \bibinfo
  {author} {\bibfnamefont {N.~A.}\ \bibnamefont {Spaldin}}, \bibinfo {author}
  {\bibfnamefont {P.}~\bibnamefont {Gambardella}},\ and\ \bibinfo {author}
  {\bibfnamefont {S.}~\bibnamefont {Stepanow}},\ }\bibfield  {journal}
  {\bibinfo  {journal} {Science Advances}\ }\textbf {\bibinfo {volume} {6}},\
  \href {https://doi.org/10.1126/SCIADV.ABC5511} {10.1126/SCIADV.ABC5511}
  (\bibinfo {year} {2020}),\ \Eprint {https://arxiv.org/abs/2005.07455}
  {arXiv:2005.07455} \BibitemShut {NoStop}%
\bibitem [{\citenamefont {Delgado}\ and\ \citenamefont
  {Lorente}(2021)}]{Delgado2021}%
  \BibitemOpen
  \bibfield  {author} {\bibinfo {author} {\bibfnamefont {F.}~\bibnamefont
  {Delgado}}\ and\ \bibinfo {author} {\bibfnamefont {N.}~\bibnamefont
  {Lorente}},\ }\href {https://doi.org/10.1016/J.PROGSURF.2021.100625}
  {\bibfield  {journal} {\bibinfo  {journal} {Progress in Surface Science}\
  }\textbf {\bibinfo {volume} {96}},\ \bibinfo {pages} {100625} (\bibinfo
  {year} {2021})}\BibitemShut {NoStop}%
\bibitem [{\citenamefont {Lounis}\ \emph {et~al.}(2010)\citenamefont {Lounis},
  \citenamefont {Costa}, \citenamefont {Muniz},\ and\ \citenamefont
  {Mills}}]{Lounis2010}%
  \BibitemOpen
  \bibfield  {author} {\bibinfo {author} {\bibfnamefont {S.}~\bibnamefont
  {Lounis}}, \bibinfo {author} {\bibfnamefont {A.~T.}\ \bibnamefont {Costa}},
  \bibinfo {author} {\bibfnamefont {R.~B.}\ \bibnamefont {Muniz}},\ and\
  \bibinfo {author} {\bibfnamefont {D.~L.}\ \bibnamefont {Mills}},\ }\href
  {https://doi.org/10.1103/PhysRevLett.105.187205} {\bibfield  {journal}
  {\bibinfo  {journal} {Physical Review Letters}\ }\textbf {\bibinfo {volume}
  {105}},\ \bibinfo {pages} {187205} (\bibinfo {year} {2010})}\BibitemShut
  {NoStop}%
\bibitem [{\citenamefont {Khajetoorians}\ \emph {et~al.}(2011)\citenamefont
  {Khajetoorians}, \citenamefont {Lounis}, \citenamefont {Chilian},
  \citenamefont {Costa}, \citenamefont {Zhou}, \citenamefont {Mills},
  \citenamefont {Wiebe},\ and\ \citenamefont
  {Wiesendanger}}]{Khajetoorians2011}%
  \BibitemOpen
  \bibfield  {author} {\bibinfo {author} {\bibfnamefont {A.~A.}\ \bibnamefont
  {Khajetoorians}}, \bibinfo {author} {\bibfnamefont {S.}~\bibnamefont
  {Lounis}}, \bibinfo {author} {\bibfnamefont {B.}~\bibnamefont {Chilian}},
  \bibinfo {author} {\bibfnamefont {A.~T.}\ \bibnamefont {Costa}}, \bibinfo
  {author} {\bibfnamefont {L.}~\bibnamefont {Zhou}}, \bibinfo {author}
  {\bibfnamefont {D.~L.}\ \bibnamefont {Mills}}, \bibinfo {author}
  {\bibfnamefont {J.}~\bibnamefont {Wiebe}},\ and\ \bibinfo {author}
  {\bibfnamefont {R.}~\bibnamefont {Wiesendanger}},\ }\href
  {https://doi.org/10.1103/PhysRevLett.106.037205} {\bibfield  {journal}
  {\bibinfo  {journal} {Physical Review Letters}\ }\textbf {\bibinfo {volume}
  {106}},\ \bibinfo {pages} {037205} (\bibinfo {year} {2011})}\BibitemShut
  {NoStop}%
\bibitem [{\citenamefont {Yang}\ \emph {et~al.}(2011)\citenamefont {Yang},
  \citenamefont {Chshiev}, \citenamefont {Dieny}, \citenamefont {Lee},
  \citenamefont {Manchon},\ and\ \citenamefont {Shin}}]{Yang2011}%
  \BibitemOpen
  \bibfield  {author} {\bibinfo {author} {\bibfnamefont {H.~X.}\ \bibnamefont
  {Yang}}, \bibinfo {author} {\bibfnamefont {M.}~\bibnamefont {Chshiev}},
  \bibinfo {author} {\bibfnamefont {B.}~\bibnamefont {Dieny}}, \bibinfo
  {author} {\bibfnamefont {J.~H.}\ \bibnamefont {Lee}}, \bibinfo {author}
  {\bibfnamefont {A.}~\bibnamefont {Manchon}},\ and\ \bibinfo {author}
  {\bibfnamefont {K.~H.}\ \bibnamefont {Shin}},\ }\href
  {https://doi.org/10.1103/PhysRevB.84.054401} {\bibfield  {journal} {\bibinfo
  {journal} {Physical Review B}\ }\textbf {\bibinfo {volume} {84}},\ \bibinfo
  {pages} {054401} (\bibinfo {year} {2011})}\BibitemShut {NoStop}%
\bibitem [{\citenamefont {Lounis}\ \emph {et~al.}(2015)\citenamefont {Lounis},
  \citenamefont {{dos Santos Dias}},\ and\ \citenamefont
  {Schweflinghaus}}]{Lounis2015}%
  \BibitemOpen
  \bibfield  {author} {\bibinfo {author} {\bibfnamefont {S.}~\bibnamefont
  {Lounis}}, \bibinfo {author} {\bibfnamefont {M.}~\bibnamefont {{dos Santos
  Dias}}},\ and\ \bibinfo {author} {\bibfnamefont {B.}~\bibnamefont
  {Schweflinghaus}},\ }\href {https://doi.org/10.1103/PhysRevB.91.104420}
  {\bibfield  {journal} {\bibinfo  {journal} {Physical Review B}\ }\textbf
  {\bibinfo {volume} {91}},\ \bibinfo {pages} {104420} (\bibinfo {year}
  {2015})}\BibitemShut {NoStop}%
\bibitem [{\citenamefont {Ferr{\'{o}}n}\ \emph {et~al.}(2015)\citenamefont
  {Ferr{\'{o}}n}, \citenamefont {Lado},\ and\ \citenamefont
  {Fern{\'{a}}ndez-Rossier}}]{Ferron2015a}%
  \BibitemOpen
  \bibfield  {author} {\bibinfo {author} {\bibfnamefont {A.}~\bibnamefont
  {Ferr{\'{o}}n}}, \bibinfo {author} {\bibfnamefont {J.~L.}\ \bibnamefont
  {Lado}},\ and\ \bibinfo {author} {\bibfnamefont {J.}~\bibnamefont
  {Fern{\'{a}}ndez-Rossier}},\ }\href
  {https://doi.org/10.1103/PhysRevB.92.174407} {\bibfield  {journal} {\bibinfo
  {journal} {Physical Review B}\ }\textbf {\bibinfo {volume} {92}},\ \bibinfo
  {pages} {174407} (\bibinfo {year} {2015})}\BibitemShut {NoStop}%
\bibitem [{\citenamefont {Khajetoorians}\ \emph {et~al.}(2016)\citenamefont
  {Khajetoorians}, \citenamefont {Steinbrecher}, \citenamefont {Ternes},
  \citenamefont {Bouhassoune}, \citenamefont {{dos Santos Dias}}, \citenamefont
  {Lounis}, \citenamefont {Wiebe},\ and\ \citenamefont
  {Wiesendanger}}]{Khajetoorians2016}%
  \BibitemOpen
  \bibfield  {author} {\bibinfo {author} {\bibfnamefont {A.~A.}\ \bibnamefont
  {Khajetoorians}}, \bibinfo {author} {\bibfnamefont {M.}~\bibnamefont
  {Steinbrecher}}, \bibinfo {author} {\bibfnamefont {M.}~\bibnamefont
  {Ternes}}, \bibinfo {author} {\bibfnamefont {M.}~\bibnamefont {Bouhassoune}},
  \bibinfo {author} {\bibfnamefont {M.}~\bibnamefont {{dos Santos Dias}}},
  \bibinfo {author} {\bibfnamefont {S.}~\bibnamefont {Lounis}}, \bibinfo
  {author} {\bibfnamefont {J.}~\bibnamefont {Wiebe}},\ and\ \bibinfo {author}
  {\bibfnamefont {R.}~\bibnamefont {Wiesendanger}},\ }\href
  {https://doi.org/10.1038/ncomms10620} {\bibfield  {journal} {\bibinfo
  {journal} {Nature Communications}\ }\textbf {\bibinfo {volume} {7}},\
  \bibinfo {pages} {10620} (\bibinfo {year} {2016})}\BibitemShut {NoStop}%
\bibitem [{\citenamefont {Iba{\~{n}}ez-Azpiroz}\ \emph
  {et~al.}(2016)\citenamefont {Iba{\~{n}}ez-Azpiroz}, \citenamefont {{dos
  Santos Dias}}, \citenamefont {Bl{\"{u}}gel},\ and\ \citenamefont
  {Lounis}}]{Ibanez-Azpiroz2016}%
  \BibitemOpen
  \bibfield  {author} {\bibinfo {author} {\bibfnamefont {J.}~\bibnamefont
  {Iba{\~{n}}ez-Azpiroz}}, \bibinfo {author} {\bibfnamefont {M.}~\bibnamefont
  {{dos Santos Dias}}}, \bibinfo {author} {\bibfnamefont {S.}~\bibnamefont
  {Bl{\"{u}}gel}},\ and\ \bibinfo {author} {\bibfnamefont {S.}~\bibnamefont
  {Lounis}},\ }\href {https://doi.org/10.1021/acs.nanolett.6b01344} {\bibfield
  {journal} {\bibinfo  {journal} {Nano Letters}\ }\textbf {\bibinfo {volume}
  {16}},\ \bibinfo {pages} {4305} (\bibinfo {year} {2016})}\BibitemShut
  {NoStop}%
\bibitem [{\citenamefont {Hermenau}\ \emph {et~al.}(2017)\citenamefont
  {Hermenau}, \citenamefont {Iba{\~{n}}ez-Azpiroz}, \citenamefont
  {H{\"{u}}bner}, \citenamefont {Sonntag}, \citenamefont {Baxevanis},
  \citenamefont {Ton}, \citenamefont {Steinbrecher}, \citenamefont
  {Khajetoorians}, \citenamefont {{dos Santos Dias}}, \citenamefont
  {Bl{\"{u}}gel}, \citenamefont {Wiesendanger}, \citenamefont {Lounis},\ and\
  \citenamefont {Wiebe}}]{Hermenau2017}%
  \BibitemOpen
  \bibfield  {author} {\bibinfo {author} {\bibfnamefont {J.}~\bibnamefont
  {Hermenau}}, \bibinfo {author} {\bibfnamefont {J.}~\bibnamefont
  {Iba{\~{n}}ez-Azpiroz}}, \bibinfo {author} {\bibfnamefont {C.}~\bibnamefont
  {H{\"{u}}bner}}, \bibinfo {author} {\bibfnamefont {A.}~\bibnamefont
  {Sonntag}}, \bibinfo {author} {\bibfnamefont {B.}~\bibnamefont {Baxevanis}},
  \bibinfo {author} {\bibfnamefont {K.~T.}\ \bibnamefont {Ton}}, \bibinfo
  {author} {\bibfnamefont {M.}~\bibnamefont {Steinbrecher}}, \bibinfo {author}
  {\bibfnamefont {A.~A.}\ \bibnamefont {Khajetoorians}}, \bibinfo {author}
  {\bibfnamefont {M.}~\bibnamefont {{dos Santos Dias}}}, \bibinfo {author}
  {\bibfnamefont {S.}~\bibnamefont {Bl{\"{u}}gel}}, \bibinfo {author}
  {\bibfnamefont {R.}~\bibnamefont {Wiesendanger}}, \bibinfo {author}
  {\bibfnamefont {S.}~\bibnamefont {Lounis}},\ and\ \bibinfo {author}
  {\bibfnamefont {J.}~\bibnamefont {Wiebe}},\ }\href
  {https://doi.org/10.1038/s41467-017-00506-7} {\bibfield  {journal} {\bibinfo
  {journal} {Nature Communications}\ }\textbf {\bibinfo {volume} {8}},\
  \bibinfo {pages} {642} (\bibinfo {year} {2017})}\BibitemShut {NoStop}%
\bibitem [{\citenamefont {Iba{\~{n}}ez-Azpiroz}\ \emph
  {et~al.}(2017{\natexlab{a}})\citenamefont {Iba{\~{n}}ez-Azpiroz},
  \citenamefont {{dos Santos Dias}}, \citenamefont {Schweflinghaus},
  \citenamefont {Bl{\"{u}}gel},\ and\ \citenamefont
  {Lounis}}]{Ibanez-Azpiroz2017}%
  \BibitemOpen
  \bibfield  {author} {\bibinfo {author} {\bibfnamefont {J.}~\bibnamefont
  {Iba{\~{n}}ez-Azpiroz}}, \bibinfo {author} {\bibfnamefont {M.}~\bibnamefont
  {{dos Santos Dias}}}, \bibinfo {author} {\bibfnamefont {B.}~\bibnamefont
  {Schweflinghaus}}, \bibinfo {author} {\bibfnamefont {S.}~\bibnamefont
  {Bl{\"{u}}gel}},\ and\ \bibinfo {author} {\bibfnamefont {S.}~\bibnamefont
  {Lounis}},\ }\href {https://doi.org/10.1103/PhysRevLett.119.017203}
  {\bibfield  {journal} {\bibinfo  {journal} {Physical Review Letters}\
  }\textbf {\bibinfo {volume} {119}},\ \bibinfo {pages} {017203} (\bibinfo
  {year} {2017}{\natexlab{a}})}\BibitemShut {NoStop}%
\bibitem [{\citenamefont {Iba{\~{n}}ez-Azpiroz}\ \emph
  {et~al.}(2017{\natexlab{b}})\citenamefont {Iba{\~{n}}ez-Azpiroz},
  \citenamefont {{dos Santos Dias}}, \citenamefont {Bl{\"{u}}gel},\ and\
  \citenamefont {Lounis}}]{Ibanez-Azpiroz2017a}%
  \BibitemOpen
  \bibfield  {author} {\bibinfo {author} {\bibfnamefont {J.}~\bibnamefont
  {Iba{\~{n}}ez-Azpiroz}}, \bibinfo {author} {\bibfnamefont {M.}~\bibnamefont
  {{dos Santos Dias}}}, \bibinfo {author} {\bibfnamefont {S.}~\bibnamefont
  {Bl{\"{u}}gel}},\ and\ \bibinfo {author} {\bibfnamefont {S.}~\bibnamefont
  {Lounis}},\ }\href {https://doi.org/10.1103/PhysRevB.96.144410} {\bibfield
  {journal} {\bibinfo  {journal} {Physical Review B}\ }\textbf {\bibinfo
  {volume} {96}},\ \bibinfo {pages} {144410} (\bibinfo {year}
  {2017}{\natexlab{b}})}\BibitemShut {NoStop}%
\bibitem [{\citenamefont {Wolf}\ \emph {et~al.}(2020)\citenamefont {Wolf},
  \citenamefont {Delgado}, \citenamefont {Reina},\ and\ \citenamefont
  {Lorente}}]{Wolf2020}%
  \BibitemOpen
  \bibfield  {author} {\bibinfo {author} {\bibfnamefont {C.}~\bibnamefont
  {Wolf}}, \bibinfo {author} {\bibfnamefont {F.}~\bibnamefont {Delgado}},
  \bibinfo {author} {\bibfnamefont {J.}~\bibnamefont {Reina}},\ and\ \bibinfo
  {author} {\bibfnamefont {N.}~\bibnamefont {Lorente}},\ }\href
  {https://doi.org/10.1021/acs.jpca.9b10749} {\bibfield  {journal} {\bibinfo
  {journal} {The Journal of Physical Chemistry A}\ }\textbf {\bibinfo {volume}
  {124}},\ \bibinfo {pages} {2318} (\bibinfo {year} {2020})}\BibitemShut
  {NoStop}%
\bibitem [{\citenamefont {Hirjibehedin}\ \emph {et~al.}(2007)\citenamefont
  {Hirjibehedin}, \citenamefont {Lin}, \citenamefont {Otte}, \citenamefont
  {Ternes}, \citenamefont {Lutz}, \citenamefont {Jones},\ and\ \citenamefont
  {Heinrich}}]{Hirjibehedin2007}%
  \BibitemOpen
  \bibfield  {author} {\bibinfo {author} {\bibfnamefont {C.~F.}\ \bibnamefont
  {Hirjibehedin}}, \bibinfo {author} {\bibfnamefont {C.-Y.}\ \bibnamefont
  {Lin}}, \bibinfo {author} {\bibfnamefont {A.~F.}\ \bibnamefont {Otte}},
  \bibinfo {author} {\bibfnamefont {M.}~\bibnamefont {Ternes}}, \bibinfo
  {author} {\bibfnamefont {C.~P.}\ \bibnamefont {Lutz}}, \bibinfo {author}
  {\bibfnamefont {B.~A.}\ \bibnamefont {Jones}},\ and\ \bibinfo {author}
  {\bibfnamefont {A.~J.}\ \bibnamefont {Heinrich}},\ }\href
  {https://doi.org/10.1126/science.1146110} {\bibfield  {journal} {\bibinfo
  {journal} {Science}\ }\textbf {\bibinfo {volume} {317}},\ \bibinfo {pages}
  {1199} (\bibinfo {year} {2007})}\BibitemShut {NoStop}%
\bibitem [{\citenamefont {Lunghi}\ \emph
  {et~al.}(2017{\natexlab{a}})\citenamefont {Lunghi}, \citenamefont {Totti},
  \citenamefont {Sessoli},\ and\ \citenamefont {Sanvito}}]{Lunghi2017}%
  \BibitemOpen
  \bibfield  {author} {\bibinfo {author} {\bibfnamefont {A.}~\bibnamefont
  {Lunghi}}, \bibinfo {author} {\bibfnamefont {F.}~\bibnamefont {Totti}},
  \bibinfo {author} {\bibfnamefont {R.}~\bibnamefont {Sessoli}},\ and\ \bibinfo
  {author} {\bibfnamefont {S.}~\bibnamefont {Sanvito}},\ }\href
  {https://doi.org/10.1038/ncomms14620} {\bibfield  {journal} {\bibinfo
  {journal} {Nature Communications}\ }\textbf {\bibinfo {volume} {8}},\
  \bibinfo {pages} {14620} (\bibinfo {year} {2017}{\natexlab{a}})}\BibitemShut
  {NoStop}%
\bibitem [{\citenamefont {Lunghi}\ \emph
  {et~al.}(2017{\natexlab{b}})\citenamefont {Lunghi}, \citenamefont {Totti},
  \citenamefont {Sanvito},\ and\ \citenamefont {Sessoli}}]{Lunghi2017a}%
  \BibitemOpen
  \bibfield  {author} {\bibinfo {author} {\bibfnamefont {A.}~\bibnamefont
  {Lunghi}}, \bibinfo {author} {\bibfnamefont {F.}~\bibnamefont {Totti}},
  \bibinfo {author} {\bibfnamefont {S.}~\bibnamefont {Sanvito}},\ and\ \bibinfo
  {author} {\bibfnamefont {R.}~\bibnamefont {Sessoli}},\ }\href
  {https://doi.org/10.1039/C7SC02832F} {\bibfield  {journal} {\bibinfo
  {journal} {Chemical Science}\ }\textbf {\bibinfo {volume} {8}},\ \bibinfo
  {pages} {6051} (\bibinfo {year} {2017}{\natexlab{b}})}\BibitemShut {NoStop}%
\bibitem [{\citenamefont {Escalera-Moreno}\ \emph {et~al.}(2017)\citenamefont
  {Escalera-Moreno}, \citenamefont {Suaud}, \citenamefont {Gaita-Ari{\~{n}}o},\
  and\ \citenamefont {Coronado}}]{Escalera-Moreno2017}%
  \BibitemOpen
  \bibfield  {author} {\bibinfo {author} {\bibfnamefont {L.}~\bibnamefont
  {Escalera-Moreno}}, \bibinfo {author} {\bibfnamefont {N.}~\bibnamefont
  {Suaud}}, \bibinfo {author} {\bibfnamefont {A.}~\bibnamefont
  {Gaita-Ari{\~{n}}o}},\ and\ \bibinfo {author} {\bibfnamefont
  {E.}~\bibnamefont {Coronado}},\ }\href
  {https://doi.org/10.1021/acs.jpclett.7b00479} {\bibfield  {journal} {\bibinfo
   {journal} {The Journal of Physical Chemistry Letters}\ }\textbf {\bibinfo
  {volume} {8}},\ \bibinfo {pages} {1695} (\bibinfo {year} {2017})}\BibitemShut
  {NoStop}%
\bibitem [{\citenamefont {Escalera-Moreno}\ \emph {et~al.}(2018)\citenamefont
  {Escalera-Moreno}, \citenamefont {Baldov{\'{i}}}, \citenamefont
  {Gaita-Ari{\~{n}}o},\ and\ \citenamefont {Coronado}}]{Escalera-Moreno2018}%
  \BibitemOpen
  \bibfield  {author} {\bibinfo {author} {\bibfnamefont {L.}~\bibnamefont
  {Escalera-Moreno}}, \bibinfo {author} {\bibfnamefont {J.~J.}\ \bibnamefont
  {Baldov{\'{i}}}}, \bibinfo {author} {\bibfnamefont {A.}~\bibnamefont
  {Gaita-Ari{\~{n}}o}},\ and\ \bibinfo {author} {\bibfnamefont
  {E.}~\bibnamefont {Coronado}},\ }\href {https://doi.org/10.1039/C7SC05464E}
  {\bibfield  {journal} {\bibinfo  {journal} {Chemical Science}\ }\textbf
  {\bibinfo {volume} {9}},\ \bibinfo {pages} {3265} (\bibinfo {year}
  {2018})}\BibitemShut {NoStop}%
\bibitem [{\citenamefont {Albino}\ \emph {et~al.}(2019)\citenamefont {Albino},
  \citenamefont {Benci}, \citenamefont {Tesi}, \citenamefont {Atzori},
  \citenamefont {Torre}, \citenamefont {Sanvito}, \citenamefont {Sessoli},\
  and\ \citenamefont {Lunghi}}]{Albino2019}%
  \BibitemOpen
  \bibfield  {author} {\bibinfo {author} {\bibfnamefont {A.}~\bibnamefont
  {Albino}}, \bibinfo {author} {\bibfnamefont {S.}~\bibnamefont {Benci}},
  \bibinfo {author} {\bibfnamefont {L.}~\bibnamefont {Tesi}}, \bibinfo {author}
  {\bibfnamefont {M.}~\bibnamefont {Atzori}}, \bibinfo {author} {\bibfnamefont
  {R.}~\bibnamefont {Torre}}, \bibinfo {author} {\bibfnamefont
  {S.}~\bibnamefont {Sanvito}}, \bibinfo {author} {\bibfnamefont
  {R.}~\bibnamefont {Sessoli}},\ and\ \bibinfo {author} {\bibfnamefont
  {A.}~\bibnamefont {Lunghi}},\ }\href
  {https://doi.org/10.1021/acs.inorgchem.9b01407} {\bibfield  {journal}
  {\bibinfo  {journal} {Inorganic Chemistry}\ }\textbf {\bibinfo {volume}
  {58}},\ \bibinfo {pages} {10260} (\bibinfo {year} {2019})},\ \Eprint
  {https://arxiv.org/abs/1904.04922} {arXiv:1904.04922} \BibitemShut {NoStop}%
\bibitem [{\citenamefont {Lunghi}\ and\ \citenamefont
  {Sanvito}(2019)}]{Lunghi2019}%
  \BibitemOpen
  \bibfield  {author} {\bibinfo {author} {\bibfnamefont {A.}~\bibnamefont
  {Lunghi}}\ and\ \bibinfo {author} {\bibfnamefont {S.}~\bibnamefont
  {Sanvito}},\ }\href {https://doi.org/10.1126/sciadv.aax7163} {\bibfield
  {journal} {\bibinfo  {journal} {Science Advances}\ }\textbf {\bibinfo
  {volume} {5}},\ \bibinfo {pages} {eaax7163} (\bibinfo {year}
  {2019})}\BibitemShut {NoStop}%
\bibitem [{\citenamefont {Lunghi}\ and\ \citenamefont
  {Sanvito}(2020)}]{Lunghi2020}%
  \BibitemOpen
  \bibfield  {author} {\bibinfo {author} {\bibfnamefont {A.}~\bibnamefont
  {Lunghi}}\ and\ \bibinfo {author} {\bibfnamefont {S.}~\bibnamefont
  {Sanvito}},\ }\href {https://doi.org/10.1063/5.0017118} {\bibfield  {journal}
  {\bibinfo  {journal} {The Journal of Chemical Physics}\ }\textbf {\bibinfo
  {volume} {153}},\ \bibinfo {pages} {174113} (\bibinfo {year}
  {2020})}\BibitemShut {NoStop}%
\bibitem [{\citenamefont {Briganti}\ \emph {et~al.}(2021)\citenamefont
  {Briganti}, \citenamefont {Santanni}, \citenamefont {Tesi}, \citenamefont
  {Totti}, \citenamefont {Sessoli},\ and\ \citenamefont
  {Lunghi}}]{Briganti2021}%
  \BibitemOpen
  \bibfield  {author} {\bibinfo {author} {\bibfnamefont {M.}~\bibnamefont
  {Briganti}}, \bibinfo {author} {\bibfnamefont {F.}~\bibnamefont {Santanni}},
  \bibinfo {author} {\bibfnamefont {L.}~\bibnamefont {Tesi}}, \bibinfo {author}
  {\bibfnamefont {F.}~\bibnamefont {Totti}}, \bibinfo {author} {\bibfnamefont
  {R.}~\bibnamefont {Sessoli}},\ and\ \bibinfo {author} {\bibfnamefont
  {A.}~\bibnamefont {Lunghi}},\ }\href {https://doi.org/10.1021/jacs.1c05068}
  {\bibfield  {journal} {\bibinfo  {journal} {Journal of the American Chemical
  Society}\ }\textbf {\bibinfo {volume} {143}},\ \bibinfo {pages} {13633}
  (\bibinfo {year} {2021})},\ \Eprint {https://arxiv.org/abs/2105.06953}
  {arXiv:2105.06953} \BibitemShut {NoStop}%
\bibitem [{\citenamefont {Donati}\ \emph {et~al.}(2020)\citenamefont {Donati},
  \citenamefont {Rusponi}, \citenamefont {Stepanow}, \citenamefont
  {Persichetti}, \citenamefont {Singha}, \citenamefont {Juraschek},
  \citenamefont {W{\"{a}}ckerlin}, \citenamefont {Baltic}, \citenamefont
  {Pivetta}, \citenamefont {Diller}, \citenamefont {Nistor}, \citenamefont
  {Dreiser}, \citenamefont {Kummer}, \citenamefont {Velez-Fort}, \citenamefont
  {Spaldin}, \citenamefont {Brune},\ and\ \citenamefont
  {Gambardella}}]{Donati2020}%
  \BibitemOpen
  \bibfield  {author} {\bibinfo {author} {\bibfnamefont {F.}~\bibnamefont
  {Donati}}, \bibinfo {author} {\bibfnamefont {S.}~\bibnamefont {Rusponi}},
  \bibinfo {author} {\bibfnamefont {S.}~\bibnamefont {Stepanow}}, \bibinfo
  {author} {\bibfnamefont {L.}~\bibnamefont {Persichetti}}, \bibinfo {author}
  {\bibfnamefont {A.}~\bibnamefont {Singha}}, \bibinfo {author} {\bibfnamefont
  {D.~M.}\ \bibnamefont {Juraschek}}, \bibinfo {author} {\bibfnamefont
  {C.}~\bibnamefont {W{\"{a}}ckerlin}}, \bibinfo {author} {\bibfnamefont
  {R.}~\bibnamefont {Baltic}}, \bibinfo {author} {\bibfnamefont
  {M.}~\bibnamefont {Pivetta}}, \bibinfo {author} {\bibfnamefont
  {K.}~\bibnamefont {Diller}}, \bibinfo {author} {\bibfnamefont
  {C.}~\bibnamefont {Nistor}}, \bibinfo {author} {\bibfnamefont
  {J.}~\bibnamefont {Dreiser}}, \bibinfo {author} {\bibfnamefont
  {K.}~\bibnamefont {Kummer}}, \bibinfo {author} {\bibfnamefont
  {E.}~\bibnamefont {Velez-Fort}}, \bibinfo {author} {\bibfnamefont {N.~A.}\
  \bibnamefont {Spaldin}}, \bibinfo {author} {\bibfnamefont {H.}~\bibnamefont
  {Brune}},\ and\ \bibinfo {author} {\bibfnamefont {P.}~\bibnamefont
  {Gambardella}},\ }\href {https://doi.org/10.1103/PhysRevLett.124.077204}
  {\bibfield  {journal} {\bibinfo  {journal} {Physical Review Letters}\
  }\textbf {\bibinfo {volume} {124}},\ \bibinfo {pages} {077204} (\bibinfo
  {year} {2020})}\BibitemShut {NoStop}%
\bibitem [{\citenamefont {Garai-Marin}\ \emph {et~al.}(2021)\citenamefont
  {Garai-Marin}, \citenamefont {Iba{\~{n}}ez-Azpiroz}, \citenamefont
  {Garcia-Goiricelaya}, \citenamefont {Gurtubay},\ and\ \citenamefont
  {Eiguren}}]{Garai-Marin2021}%
  \BibitemOpen
  \bibfield  {author} {\bibinfo {author} {\bibfnamefont {H.}~\bibnamefont
  {Garai-Marin}}, \bibinfo {author} {\bibfnamefont {J.}~\bibnamefont
  {Iba{\~{n}}ez-Azpiroz}}, \bibinfo {author} {\bibfnamefont {P.}~\bibnamefont
  {Garcia-Goiricelaya}}, \bibinfo {author} {\bibfnamefont {I.~G.}\ \bibnamefont
  {Gurtubay}},\ and\ \bibinfo {author} {\bibfnamefont {A.}~\bibnamefont
  {Eiguren}},\ }\href {https://doi.org/10.1103/PhysRevB.104.195422} {\bibfield
  {journal} {\bibinfo  {journal} {Physical Review B}\ }\textbf {\bibinfo
  {volume} {104}},\ \bibinfo {pages} {195422} (\bibinfo {year}
  {2021})}\BibitemShut {NoStop}%
\bibitem [{\citenamefont {Soler}\ \emph {et~al.}(2002)\citenamefont {Soler},
  \citenamefont {Artacho}, \citenamefont {Gale}, \citenamefont {Garc{\'{i}}a},
  \citenamefont {Junquera}, \citenamefont {Ordej{\'{o}}n},\ and\ \citenamefont
  {S{\'{a}}nchez-Portal}}]{Soler2002}%
  \BibitemOpen
  \bibfield  {author} {\bibinfo {author} {\bibfnamefont {J.~M.}\ \bibnamefont
  {Soler}}, \bibinfo {author} {\bibfnamefont {E.}~\bibnamefont {Artacho}},
  \bibinfo {author} {\bibfnamefont {J.~D.}\ \bibnamefont {Gale}}, \bibinfo
  {author} {\bibfnamefont {A.}~\bibnamefont {Garc{\'{i}}a}}, \bibinfo {author}
  {\bibfnamefont {J.}~\bibnamefont {Junquera}}, \bibinfo {author}
  {\bibfnamefont {P.}~\bibnamefont {Ordej{\'{o}}n}},\ and\ \bibinfo {author}
  {\bibfnamefont {D.}~\bibnamefont {S{\'{a}}nchez-Portal}},\ }\href
  {https://doi.org/10.1088/0953-8984/14/11/302} {\bibfield  {journal} {\bibinfo
   {journal} {Journal of Physics: Condensed Matter}\ }\textbf {\bibinfo
  {volume} {14}},\ \bibinfo {pages} {2745} (\bibinfo {year}
  {2002})}\BibitemShut {NoStop}%
\bibitem [{\citenamefont {Cuadrado}\ and\ \citenamefont {{I.
  Cerd{\'{a}}}}(2012)}]{Cuadrado2012}%
  \BibitemOpen
  \bibfield  {author} {\bibinfo {author} {\bibfnamefont {R.}~\bibnamefont
  {Cuadrado}}\ and\ \bibinfo {author} {\bibfnamefont {J.}~\bibnamefont {{I.
  Cerd{\'{a}}}}},\ }\href {https://doi.org/10.1088/0953-8984/24/8/086005}
  {\bibfield  {journal} {\bibinfo  {journal} {Journal of Physics: Condensed
  Matter}\ }\textbf {\bibinfo {volume} {24}},\ \bibinfo {pages} {086005}
  (\bibinfo {year} {2012})}\BibitemShut {NoStop}%
\bibitem [{Note1()}]{Note1}%
  \BibitemOpen
  \bibinfo {note} {See Supplemental Material at [URL] , which includes
  Refs.~\cite {Perdew1996,Kleinman1982,Breuer2007}, for a detailed information
  of the DFT calculations, a list of Stevens operators and the second
  quantization expression of the basis states, together with the derivation of
  the rate equation.}\BibitemShut {Stop}%
\bibitem [{\citenamefont {Grimvall}(1983)}]{Grimvall1981}%
  \BibitemOpen
  \bibfield  {author} {\bibinfo {author} {\bibfnamefont {G.}~\bibnamefont
  {Grimvall}},\ }\href {https://doi.org/10.1002/bbpc.19830870521} {\emph
  {\bibinfo {title} {{The Electron-Phonon Interaction in Metals}}}},\ Series of
  monographs on selected topics in solid state physics\ (\bibinfo  {publisher}
  {North Holland Publishing Company},\ \bibinfo {address} {Amsterdam, New York,
  Oxford},\ \bibinfo {year} {1983})\BibitemShut {NoStop}%
\bibitem [{\citenamefont {Mahan}(2000)}]{Mahan2000}%
  \BibitemOpen
  \bibfield  {author} {\bibinfo {author} {\bibfnamefont {G.~D.}\ \bibnamefont
  {Mahan}},\ }\href {https://doi.org/10.1007/978-1-4757-5714-9} {\emph
  {\bibinfo {title} {{Many-Particle Physics}}}},\ \bibinfo {edition} {3rd}\
  ed.\ (\bibinfo  {publisher} {Springer New York, NY},\ \bibinfo {year}
  {2000})\BibitemShut {NoStop}%
\bibitem [{\citenamefont {Giustino}(2017)}]{Giustino2017}%
  \BibitemOpen
  \bibfield  {author} {\bibinfo {author} {\bibfnamefont {F.}~\bibnamefont
  {Giustino}},\ }\href {https://doi.org/10.1103/RevModPhys.89.015003}
  {\bibfield  {journal} {\bibinfo  {journal} {Reviews of Modern Physics}\
  }\textbf {\bibinfo {volume} {89}},\ \bibinfo {pages} {015003} (\bibinfo
  {year} {2017})}\BibitemShut {NoStop}%
\bibitem [{\citenamefont {Baumann}(2015)}]{Baumann2015a}%
  \BibitemOpen
  \bibfield  {author} {\bibinfo {author} {\bibfnamefont {S.}~\bibnamefont
  {Baumann}},\ }\href {https://doi.org/10.5451/unibas-006489486} {Ph.D.
  thesis},\ \bibinfo  {school} {University of Basel} (\bibinfo {year}
  {2015})\BibitemShut {NoStop}%
\bibitem [{\citenamefont {van Weerdenburg}\ \emph {et~al.}(2021)\citenamefont
  {van Weerdenburg}, \citenamefont {Steinbrecher}, \citenamefont {van
  Mullekom}, \citenamefont {Gerritsen}, \citenamefont {von Allw{\"{o}}rden},
  \citenamefont {Natterer},\ and\ \citenamefont
  {Khajetoorians}}]{VanWeerdenburg2021}%
  \BibitemOpen
  \bibfield  {author} {\bibinfo {author} {\bibfnamefont {W.~M.~J.}\
  \bibnamefont {van Weerdenburg}}, \bibinfo {author} {\bibfnamefont
  {M.}~\bibnamefont {Steinbrecher}}, \bibinfo {author} {\bibfnamefont
  {N.~P.~E.}\ \bibnamefont {van Mullekom}}, \bibinfo {author} {\bibfnamefont
  {J.~W.}\ \bibnamefont {Gerritsen}}, \bibinfo {author} {\bibfnamefont
  {H.}~\bibnamefont {von Allw{\"{o}}rden}}, \bibinfo {author} {\bibfnamefont
  {F.~D.}\ \bibnamefont {Natterer}},\ and\ \bibinfo {author} {\bibfnamefont
  {A.~A.}\ \bibnamefont {Khajetoorians}},\ }\href
  {https://doi.org/10.1063/5.0040011} {\bibfield  {journal} {\bibinfo
  {journal} {Review of Scientific Instruments}\ }\textbf {\bibinfo {volume}
  {92}},\ \bibinfo {pages} {033906} (\bibinfo {year} {2021})}\BibitemShut
  {NoStop}%
\bibitem [{\citenamefont {Perdew}\ \emph {et~al.}(1996)\citenamefont {Perdew},
  \citenamefont {Burke},\ and\ \citenamefont {Ernzerhof}}]{Perdew1996}%
  \BibitemOpen
  \bibfield  {author} {\bibinfo {author} {\bibfnamefont {J.~P.}\ \bibnamefont
  {Perdew}}, \bibinfo {author} {\bibfnamefont {K.}~\bibnamefont {Burke}},\ and\
  \bibinfo {author} {\bibfnamefont {M.}~\bibnamefont {Ernzerhof}},\ }\href
  {https://doi.org/10.1103/PhysRevLett.77.3865} {\bibfield  {journal} {\bibinfo
   {journal} {Physical Review Letters}\ }\textbf {\bibinfo {volume} {77}},\
  \bibinfo {pages} {3865} (\bibinfo {year} {1996})}\BibitemShut {NoStop}%
\bibitem [{\citenamefont {Kleinman}\ and\ \citenamefont
  {Bylander}(1982)}]{Kleinman1982}%
  \BibitemOpen
  \bibfield  {author} {\bibinfo {author} {\bibfnamefont {L.}~\bibnamefont
  {Kleinman}}\ and\ \bibinfo {author} {\bibfnamefont {D.~M.}\ \bibnamefont
  {Bylander}},\ }\href {https://doi.org/10.1103/PhysRevLett.48.1425} {\bibfield
   {journal} {\bibinfo  {journal} {Physical Review Letters}\ }\textbf {\bibinfo
  {volume} {48}},\ \bibinfo {pages} {1425} (\bibinfo {year}
  {1982})}\BibitemShut {NoStop}%
\bibitem [{\citenamefont {Breuer}\ and\ \citenamefont
  {Petruccione}(2007)}]{Breuer2007}%
  \BibitemOpen
  \bibfield  {author} {\bibinfo {author} {\bibfnamefont {H.-P.}\ \bibnamefont
  {Breuer}}\ and\ \bibinfo {author} {\bibfnamefont {F.}~\bibnamefont
  {Petruccione}},\ }\href
  {https://doi.org/10.1093/acprof:oso/9780199213900.001.0001} {\emph {\bibinfo
  {title} {{The Theory of Open Quantum Systems}}}}\ (\bibinfo  {publisher}
  {Oxford University Press},\ \bibinfo {year} {2007})\BibitemShut {NoStop}%
\end{thebibliography}%

\clearpage
\newpage

\begin{bibunit}[apsrev4-2]

\onecolumngrid
\begin{center}
\textbf{\large Supplemental Materials: \thetitle}
\end{center}
\setcounter{equation}{0}
\setcounter{figure}{0}
\setcounter{table}{0}
\setcounter{page}{1}
\makeatletter
\renewcommand{\theequation}{S\arabic{equation}}
\renewcommand{\thefigure}{S\arabic{figure}}
\renewcommand{\bibnumfmt}[1]{[S#1]}
\renewcommand{\citenumfont}[1]{S#1}

\newcommand{\Tr}{\mathrm{Tr}}


\section{Details of DFT calculations}

We have used relativistic density functional theory (DFT) calculations to compute the electronic and vibrational structures of the Fe/MgO/Ag(100) system and calculate the electron-phonon matrix elements.
The vibrational frequencies and the potential induced by the atomic displacements were calculated using the so-called direct method, employing symmetries to reduce computational costs, as explained in Ref.~\cite{Garai-Marin2021}.
The calculations were done using the non-colinear off-site formalism for the spin-orbit coupling implemented in SIESTA \cite{Soler2002,Cuadrado2012}. The unit cell consisted of a $4\times4\times1$ super cell of a MgO/Ag(100) slab, with an iron adatom adsorbed on top of an oxygen site. The MgO/Ag(100) slab is modeled by 11 Ag layers and an overlayer of 2 to 4 monolayers of MgO in both terminations. We optimized the geometry of the system keeping fixed the positions of the inner 5 layers of silver. The generalized gradient approximation parametrized by Perdew, Burke, and Ernzerhof \cite{Perdew1996} (PBE-GGA) has been used for the exchange-correlation functional. Core electrons are represented using separable \cite{Kleinman1982} norm-conserving Pseudo-Potentials (PPs) and valence electrons are expanded using optimized basis sets for silver and oxygen, a triple-zeta plus 2 polarization orbitals for magnesium and a triple-zeta plus 3 polarization orbitals for iron. The $\Gamma$-point was used for integration of the Brillouin zone, and real space integrals were computed using a mesh cutoff of 600 Ry.
together with the Grid.CellSampling parameter to mitigate the egg-box effect on atomic forces and properly determine the soft modes of the adatom. A Fermi-Dirac distribution function for an electronic temperature of 300 K was used to compute the occupations.

\section{Stevens operators}

The Stevens operators used on this work are the following:
\begin{align}
\hat{O}_2^0(\mathbf{L})
=
3 L_z^2 - X
,\end{align}
\begin{align}
\hat{O}_4^0(\mathbf{L})
=
35 L_z^4 - (30 X-25)L_z^2 + 3 X^2 - 6 X
\end{align}
and
\begin{align}
\hat{O}_4^4(\mathbf{L})
=
\frac{1}{2}(L_+^4 + L_-^4)
.\end{align}
Where $X = L(L+1)$.

For a extended table of Stevens operators visit \url{https://easyspin.org/easyspin/documentation/stevensoperators.html}.

\section{Basis states in second quantization}

To obtain the second quantization expression of the basis states $|M_\mathrm{S},M_\mathrm{L}\rangle$ of the $^5D$ term ($S=2$ and $L=2$), we have started from the trivial maximum $M_L=2$ and $M_S=2$ state given by
\begin{equation}
|2,2\rangle=c_{2\downarrow}^\dagger c_{-2\uparrow}^\dagger c_{-1\uparrow}^\dagger c_{0\uparrow}^\dagger c_{1\uparrow}^\dagger c_{2\uparrow}^\dagger |0\rangle
,\end{equation}
where subindices denote orbital angular momentum $z$-projection and spin, respectively, and $|0\rangle$ denotes the state with no $d$ electrons. The order for the operators chosen along the work is placing spin majority operators on the right, with highest orbital angular momentum $z$-projection on the right.

The remaining states can be obtained by applying $S_-$ and $L_-$ operators. All the states are listed below:

\small
\begin{align}
\label{eq:sq-basis}
|M_L,M_S\rangle & 
\\ \nonumber
|\phantom{-}2,\phantom{-}2\rangle
& =
c_{2\downarrow}^\dagger c_{-2\uparrow}^\dagger c_{-1\uparrow}^\dagger c_{0\uparrow}^\dagger c_{1\uparrow}^\dagger c_{2\uparrow}^\dagger 
|0\rangle
\\ \nonumber
|\phantom{-}2,\phantom{-}1\rangle
& =
\frac{1}{\sqrt{4}}
\left(
c_{1\downarrow}^\dagger c_{2\downarrow}^\dagger c_{-2\uparrow}^\dagger c_{-1\uparrow}^\dagger c_{0\uparrow}^\dagger c_{2\uparrow}^\dagger
-
c_{0\downarrow}^\dagger c_{2\downarrow}^\dagger c_{-2\uparrow}^\dagger c_{-1\uparrow}^\dagger c_{1\uparrow}^\dagger c_{2\uparrow}^\dagger
+
c_{-1\downarrow}^\dagger c_{2\downarrow}^\dagger c_{-2\uparrow}^\dagger c_{0\uparrow}^\dagger c_{1\uparrow}^\dagger c_{2\uparrow}^\dagger
-
c_{-2\downarrow}^\dagger c_{2\downarrow}^\dagger c_{-1\uparrow}^\dagger c_{0\uparrow}^\dagger c_{1\uparrow}^\dagger c_{2\uparrow}^\dagger
\right)
|0\rangle
\\ \nonumber
|\phantom{-}2,\phantom{-}0\rangle
& =
\frac{1}{\sqrt{6}}
\left(
c_{0\downarrow}^\dagger c_{1\downarrow}^\dagger c_{2\downarrow}^\dagger c_{-2\uparrow}^\dagger c_{-1\uparrow}^\dagger c_{2\uparrow}^\dagger 
-
c_{-1\downarrow}^\dagger c_{1\downarrow}^\dagger c_{2\downarrow}^\dagger c_{-2\uparrow}^\dagger c_{0\uparrow}^\dagger c_{2\uparrow}^\dagger 
+
c_{-2\downarrow}^\dagger c_{1\downarrow}^\dagger c_{2\downarrow}^\dagger c_{-1\uparrow}^\dagger c_{0\uparrow}^\dagger c_{2\uparrow}^\dagger 
\right.
\\ \nonumber
& \phantom{= \frac{1}{\sqrt{6}}}
\left.
+
c_{-1\downarrow}^\dagger c_{0\downarrow}^\dagger c_{2\downarrow}^\dagger c_{-2\uparrow}^\dagger c_{1\uparrow}^\dagger c_{2\uparrow}^\dagger 
-
c_{-2\downarrow}^\dagger c_{0\downarrow}^\dagger c_{2\downarrow}^\dagger c_{-1\uparrow}^\dagger c_{1\uparrow}^\dagger c_{2\uparrow}^\dagger 
+
c_{-2\downarrow}^\dagger c_{-1\downarrow}^\dagger c_{2\downarrow}^\dagger c_{0\uparrow}^\dagger c_{1\uparrow}^\dagger c_{2\uparrow}^\dagger 
\right)
|0\rangle
\\ \nonumber
|\phantom{-}2,-1\rangle
& =
\frac{1}{\sqrt{4}}
\left( c_{-1\downarrow}^\dagger c_{0\downarrow}^\dagger c_{1\downarrow}^\dagger c_{2\downarrow}^\dagger c_{-2\uparrow}^\dagger c_{2\uparrow}^\dagger
-
c_{-2\downarrow}^\dagger c_{0\downarrow}^\dagger c_{1\downarrow}^\dagger c_{2\downarrow}^\dagger c_{-1\uparrow}^\dagger c_{2\uparrow}^\dagger
+
c_{-2\downarrow}^\dagger c_{-1\downarrow}^\dagger c_{1\downarrow}^\dagger c_{2\downarrow}^\dagger c_{0\uparrow}^\dagger c_{2\uparrow}^\dagger
-
c_{-2\downarrow}^\dagger c_{-1\downarrow}^\dagger c_{0\downarrow}^\dagger c_{2\downarrow}^\dagger c_{1\uparrow}^\dagger c_{2\uparrow}^\dagger
\right)
|0\rangle
\\ \nonumber
|\phantom{-}2,-2\rangle
& =
c_{-2\downarrow}^\dagger c_{-1\downarrow}^\dagger c_{0\downarrow}^\dagger c_{1\downarrow}^\dagger c_{2\downarrow}^\dagger c_{2\uparrow}^\dagger
|0\rangle
\\ \nonumber
|\phantom{-}1,\phantom{-}2\rangle
& =
c_{1\downarrow}^\dagger c_{-2\uparrow}^\dagger c_{-1\uparrow}^\dagger c_{0\uparrow}^\dagger c_{1\uparrow}^\dagger c_{2\uparrow}^\dagger
|0\rangle
\\ \nonumber
|\phantom{-}1,\phantom{-}1\rangle
& =
\frac{1}{\sqrt{4}}
\left(
c_{1\downarrow}^\dagger c_{2\downarrow}^\dagger c_{-2\uparrow}^\dagger c_{-1\uparrow}^\dagger c_{0\uparrow}^\dagger c_{1\uparrow}^\dagger 
-
c_{0\downarrow}^\dagger c_{1\downarrow}^\dagger c_{-2\uparrow}^\dagger c_{-1\uparrow}^\dagger c_{1\uparrow}^\dagger c_{2\uparrow}^\dagger 
+
c_{-1\downarrow}^\dagger c_{1\downarrow}^\dagger c_{-2\uparrow}^\dagger c_{0\uparrow}^\dagger c_{1\uparrow}^\dagger c_{2\uparrow}^\dagger 
-
c_{-2\downarrow}^\dagger c_{1\downarrow}^\dagger c_{-1\uparrow}^\dagger c_{0\uparrow}^\dagger c_{1\uparrow}^\dagger c_{2\uparrow}^\dagger 
\right)
|0\rangle
\\ \nonumber
|\phantom{-}1,\phantom{-}0\rangle
& =
\frac{1}{\sqrt{6}}
\left(
c_{0\downarrow}^\dagger c_{1\downarrow}^\dagger c_{2\downarrow}^\dagger c_{-2\uparrow}^\dagger c_{-1\uparrow}^\dagger c_{1\uparrow}^\dagger 
-
c_{-1\downarrow}^\dagger c_{1\downarrow}^\dagger c_{2\downarrow}^\dagger c_{-2\uparrow}^\dagger c_{0\uparrow}^\dagger c_{1\uparrow}^\dagger 
+
c_{-2\downarrow}^\dagger c_{1\downarrow}^\dagger c_{2\downarrow}^\dagger c_{-1\uparrow}^\dagger c_{0\uparrow}^\dagger c_{1\uparrow}^\dagger 
\right.
\\ \nonumber
& \phantom{= \frac{1}{\sqrt{6}}}
\left.
+
c_{-1\downarrow}^\dagger c_{0\downarrow}^\dagger c_{1\downarrow}^\dagger c_{-2\uparrow}^\dagger c_{1\uparrow}^\dagger c_{2\uparrow}^\dagger 
-
c_{-2\downarrow}^\dagger c_{0\downarrow}^\dagger c_{1\downarrow}^\dagger c_{-1\uparrow}^\dagger c_{1\uparrow}^\dagger c_{2\uparrow}^\dagger 
+
c_{-2\downarrow}^\dagger c_{-1\downarrow}^\dagger c_{1\downarrow}^\dagger c_{0\uparrow}^\dagger c_{1\uparrow}^\dagger c_{2\uparrow}^\dagger 
\right)
|0\rangle
\\ \nonumber
|\phantom{-}1,-1\rangle
& =
\frac{1}{\sqrt{4}} 
\left( 
c_{-1\downarrow}^\dagger c_{0\downarrow}^\dagger c_{1\downarrow}^\dagger c_{2\downarrow}^\dagger c_{-2\uparrow}^\dagger c_{1\uparrow}^\dagger 
-
c_{-2\downarrow}^\dagger c_{0\downarrow}^\dagger c_{1\downarrow}^\dagger c_{2\downarrow}^\dagger c_{-1\uparrow}^\dagger c_{1\uparrow}^\dagger 
+
c_{-2\downarrow}^\dagger c_{-1\downarrow}^\dagger c_{1\downarrow}^\dagger c_{2\downarrow}^\dagger c_{0\uparrow}^\dagger c_{1\uparrow}^\dagger 
-
c_{-2\downarrow}^\dagger c_{-1\downarrow}^\dagger c_{0\downarrow}^\dagger c_{1\downarrow}^\dagger c_{1\uparrow}^\dagger c_{2\uparrow}^\dagger 
\right)
|0\rangle
\\ \nonumber
|\phantom{-}1,-2\rangle
& =
c_{-2\downarrow}^\dagger c_{-1\downarrow}^\dagger c_{0\downarrow}^\dagger c_{1\downarrow}^\dagger c_{2\downarrow}^\dagger c_{1\uparrow}^\dagger 
|0\rangle
\\ \nonumber
|\phantom{-}0,\phantom{-}2\rangle
& =
c_{0\downarrow}^\dagger c_{-2\uparrow}^\dagger c_{-1\uparrow}^\dagger c_{0\uparrow}^\dagger c_{1\uparrow}^\dagger c_{2\uparrow}^\dagger 
|0\rangle
\\ \nonumber
|\phantom{-}0,\phantom{-}1\rangle
& =
\frac{1}{\sqrt{4}} 
\left( 
c_{0\downarrow}^\dagger c_{2\downarrow}^\dagger c_{-2\uparrow}^\dagger c_{-1\uparrow}^\dagger c_{0\uparrow}^\dagger c_{1\uparrow}^\dagger 
-
c_{0\downarrow}^\dagger c_{1\downarrow}^\dagger c_{-2\uparrow}^\dagger c_{-1\uparrow}^\dagger c_{0\uparrow}^\dagger c_{2\uparrow}^\dagger 
+
c_{-1\downarrow}^\dagger c_{0\downarrow}^\dagger c_{-2\uparrow}^\dagger c_{0\uparrow}^\dagger c_{1\uparrow}^\dagger c_{2\uparrow}^\dagger 
-
c_{-2\downarrow}^\dagger c_{0\downarrow}^\dagger c_{-1\uparrow}^\dagger c_{0\uparrow}^\dagger c_{1\uparrow}^\dagger c_{2\uparrow}^\dagger 
\right)
|0\rangle
\\ \nonumber
|\phantom{-}0,\phantom{-}0\rangle
& =
\frac{1}{\sqrt{6}} 
\left( 
c_{0\downarrow}^\dagger c_{1\downarrow}^\dagger c_{2\downarrow}^\dagger c_{-2\uparrow}^\dagger c_{-1\uparrow}^\dagger c_{0\uparrow}^\dagger 
-
c_{-1\downarrow}^\dagger c_{0\downarrow}^\dagger c_{2\downarrow}^\dagger c_{-2\uparrow}^\dagger c_{0\uparrow}^\dagger c_{1\uparrow}^\dagger 
+
c_{-2\downarrow}^\dagger c_{0\downarrow}^\dagger c_{2\downarrow}^\dagger c_{-1\uparrow}^\dagger c_{0\uparrow}^\dagger c_{1\uparrow}^\dagger 
\right.
\\ \nonumber
& \phantom{= \frac{1}{\sqrt{6}}}
\left.
+
c_{-1\downarrow}^\dagger c_{0\downarrow}^\dagger c_{1\downarrow}^\dagger c_{-2\uparrow}^\dagger c_{0\uparrow}^\dagger c_{2\uparrow}^\dagger 
-
c_{-2\downarrow}^\dagger c_{0\downarrow}^\dagger c_{1\downarrow}^\dagger c_{-1\uparrow}^\dagger c_{0\uparrow}^\dagger c_{2\uparrow}^\dagger 
+
c_{-2\downarrow}^\dagger c_{-1\downarrow}^\dagger c_{0\downarrow}^\dagger c_{0\uparrow}^\dagger c_{1\uparrow}^\dagger c_{2\uparrow}^\dagger 
\right)
|0\rangle
\\ \nonumber
|\phantom{-}0,-1\rangle
& =
\frac{1}{\sqrt{4}} 
\left( 
c_{-1\downarrow}^\dagger c_{0\downarrow}^\dagger c_{1\downarrow}^\dagger c_{2\downarrow}^\dagger c_{-2\uparrow}^\dagger c_{0\uparrow}^\dagger 
-
c_{-2\downarrow}^\dagger c_{0\downarrow}^\dagger c_{1\downarrow}^\dagger c_{2\downarrow}^\dagger c_{-1\uparrow}^\dagger c_{0\uparrow}^\dagger 
+
c_{-2\downarrow}^\dagger c_{-1\downarrow}^\dagger c_{0\downarrow}^\dagger c_{2\downarrow}^\dagger c_{0\uparrow}^\dagger c_{1\uparrow}^\dagger 
-
c_{-2\downarrow}^\dagger c_{-1\downarrow}^\dagger c_{0\downarrow}^\dagger c_{1\downarrow}^\dagger c_{0\uparrow}^\dagger c_{2\uparrow}^\dagger 
\right)
|0\rangle
\\ \nonumber
|\phantom{-}0,-2\rangle
& =
c_{-2\downarrow}^\dagger c_{-1\downarrow}^\dagger c_{0\downarrow}^\dagger c_{1\downarrow}^\dagger c_{2\downarrow}^\dagger c_{0\uparrow}^\dagger 
|0\rangle
\\ \nonumber
|-1,\phantom{-}2\rangle
& =
c_{-1\downarrow}^\dagger c_{-2\uparrow}^\dagger c_{-1\uparrow}^\dagger c_{0\uparrow}^\dagger c_{1\uparrow}^\dagger c_{2\uparrow}^\dagger 
|0\rangle
\\ \nonumber
|-1,\phantom{-}1\rangle
& =
\frac{1}{\sqrt{4}} 
\left( 
c_{-1\downarrow}^\dagger c_{2\downarrow}^\dagger c_{-2\uparrow}^\dagger c_{-1\uparrow}^\dagger c_{0\uparrow}^\dagger c_{1\uparrow}^\dagger 
-
c_{-1\downarrow}^\dagger c_{1\downarrow}^\dagger c_{-2\uparrow}^\dagger c_{-1\uparrow}^\dagger c_{0\uparrow}^\dagger c_{2\uparrow}^\dagger 
+
c_{-1\downarrow}^\dagger c_{0\downarrow}^\dagger c_{-2\uparrow}^\dagger c_{-1\uparrow}^\dagger c_{1\uparrow}^\dagger c_{2\uparrow}^\dagger 
-
c_{-2\downarrow}^\dagger c_{-1\downarrow}^\dagger c_{-1\uparrow}^\dagger c_{0\uparrow}^\dagger c_{1\uparrow}^\dagger c_{2\uparrow}^\dagger 
\right)
|0\rangle
\\ \nonumber
|-1,\phantom{-}0\rangle
& =
\frac{1}{\sqrt{6}} 
\left( 
c_{-1\downarrow}^\dagger c_{1\downarrow}^\dagger c_{2\downarrow}^\dagger c_{-2\uparrow}^\dagger c_{-1\uparrow}^\dagger c_{0\uparrow}^\dagger 
-
c_{-1\downarrow}^\dagger c_{0\downarrow}^\dagger c_{2\downarrow}^\dagger c_{-2\uparrow}^\dagger c_{-1\uparrow}^\dagger c_{1\uparrow}^\dagger 
+
c_{-2\downarrow}^\dagger c_{-1\downarrow}^\dagger c_{2\downarrow}^\dagger c_{-1\uparrow}^\dagger c_{0\uparrow}^\dagger c_{1\uparrow}^\dagger 
\right.
\\ \nonumber
& \phantom{= \frac{1}{\sqrt{6}}}
\left.
+
c_{-1\downarrow}^\dagger c_{0\downarrow}^\dagger c_{1\downarrow}^\dagger c_{-2\uparrow}^\dagger c_{-1\uparrow}^\dagger c_{2\uparrow}^\dagger 
-
c_{-2\downarrow}^\dagger c_{-1\downarrow}^\dagger c_{1\downarrow}^\dagger c_{-1\uparrow}^\dagger c_{0\uparrow}^\dagger c_{2\uparrow}^\dagger 
+
c_{-2\downarrow}^\dagger c_{-1\downarrow}^\dagger c_{0\downarrow}^\dagger c_{-1\uparrow}^\dagger c_{1\uparrow}^\dagger c_{2\uparrow}^\dagger 
\right)
|0\rangle
\\ \nonumber
|-1,-1\rangle
& =
\frac{1}{\sqrt{4}} 
\left( 
c_{-1\downarrow}^\dagger c_{0\downarrow}^\dagger c_{1\downarrow}^\dagger c_{2\downarrow}^\dagger c_{-2\uparrow}^\dagger c_{-1\uparrow}^\dagger 
-
c_{-2\downarrow}^\dagger c_{-1\downarrow}^\dagger c_{1\downarrow}^\dagger c_{2\downarrow}^\dagger c_{-1\uparrow}^\dagger c_{0\uparrow}^\dagger 
+
c_{-2\downarrow}^\dagger c_{-1\downarrow}^\dagger c_{0\downarrow}^\dagger c_{2\downarrow}^\dagger c_{-1\uparrow}^\dagger c_{1\uparrow}^\dagger 
-
c_{-2\downarrow}^\dagger c_{-1\downarrow}^\dagger c_{0\downarrow}^\dagger c_{1\downarrow}^\dagger c_{-1\uparrow}^\dagger c_{2\uparrow}^\dagger 
\right)
|0\rangle
\\ \nonumber
|-1,-2\rangle
& =
c_{-2\downarrow}^\dagger c_{-1\downarrow}^\dagger c_{0\downarrow}^\dagger c_{1\downarrow}^\dagger c_{2\downarrow}^\dagger c_{-1\uparrow}^\dagger 
|0\rangle
\\ \nonumber
|-2,\phantom{-}2\rangle
& =
c_{-2\downarrow}^\dagger c_{-2\uparrow}^\dagger c_{-1\uparrow}^\dagger c_{0\uparrow}^\dagger c_{1\uparrow}^\dagger c_{2\uparrow}^\dagger 
|0\rangle
\\ \nonumber
|-2,\phantom{-}1\rangle
& =
\frac{1}{\sqrt{4}} 
\left( 
c_{-2\downarrow}^\dagger c_{2\downarrow}^\dagger c_{-2\uparrow}^\dagger c_{-1\uparrow}^\dagger c_{0\uparrow}^\dagger c_{1\uparrow}^\dagger 
-
c_{-2\downarrow}^\dagger c_{1\downarrow}^\dagger c_{-2\uparrow}^\dagger c_{-1\uparrow}^\dagger c_{0\uparrow}^\dagger c_{2\uparrow}^\dagger 
+
c_{-2\downarrow}^\dagger c_{0\downarrow}^\dagger c_{-2\uparrow}^\dagger c_{-1\uparrow}^\dagger c_{1\uparrow}^\dagger c_{2\uparrow}^\dagger 
-
c_{-2\downarrow}^\dagger c_{-1\downarrow}^\dagger c_{-2\uparrow}^\dagger c_{0\uparrow}^\dagger c_{1\uparrow}^\dagger c_{2\uparrow}^\dagger 
\right)
|0\rangle
\\ \nonumber
|-2,\phantom{-}0\rangle
& =
\frac{1}{\sqrt{6}} 
\left( 
c_{-2\downarrow}^\dagger c_{1\downarrow}^\dagger c_{2\downarrow}^\dagger c_{-2\uparrow}^\dagger c_{-1\uparrow}^\dagger c_{0\uparrow}^\dagger 
-
c_{-2\downarrow}^\dagger c_{0\downarrow}^\dagger c_{2\downarrow}^\dagger c_{-2\uparrow}^\dagger c_{-1\uparrow}^\dagger c_{1\uparrow}^\dagger 
+
c_{-2\downarrow}^\dagger c_{-1\downarrow}^\dagger c_{2\downarrow}^\dagger c_{-2\uparrow}^\dagger c_{0\uparrow}^\dagger c_{1\uparrow}^\dagger
\right.
\\ \nonumber
& \phantom{= \frac{1}{\sqrt{6}}}
\left.
+
c_{-2\downarrow}^\dagger c_{0\downarrow}^\dagger c_{1\downarrow}^\dagger c_{-2\uparrow}^\dagger c_{-1\uparrow}^\dagger c_{2\uparrow}^\dagger 
-
c_{-2\downarrow}^\dagger c_{-1\downarrow}^\dagger c_{1\downarrow}^\dagger c_{-2\uparrow}^\dagger c_{0\uparrow}^\dagger c_{2\uparrow}^\dagger 
+
c_{-2\downarrow}^\dagger c_{-1\downarrow}^\dagger c_{0\downarrow}^\dagger c_{-2\uparrow}^\dagger c_{1\uparrow}^\dagger c_{2\uparrow}^\dagger 
\right)
|0\rangle
\\ \nonumber
|-2,-1\rangle
& =
\frac{1}{\sqrt{4}} 
\left( 
c_{-2\downarrow}^\dagger c_{0\downarrow}^\dagger c_{1\downarrow}^\dagger c_{2\downarrow}^\dagger c_{-2\uparrow}^\dagger c_{-1\uparrow}^\dagger
-
c_{-2\downarrow}^\dagger c_{-1\downarrow}^\dagger c_{1\downarrow}^\dagger c_{2\downarrow}^\dagger c_{-2\uparrow}^\dagger c_{0\uparrow}^\dagger 
+
c_{-2\downarrow}^\dagger c_{-1\downarrow}^\dagger c_{0\downarrow}^\dagger c_{2\downarrow}^\dagger c_{-2\uparrow}^\dagger c_{1\uparrow}^\dagger 
-
c_{-2\downarrow}^\dagger c_{-1\downarrow}^\dagger c_{0\downarrow}^\dagger c_{1\downarrow}^\dagger c_{-2\uparrow}^\dagger c_{2\uparrow}^\dagger 
\right)
|0\rangle
\\ \nonumber
|-2,-2\rangle
& =
c_{-2\downarrow}^\dagger c_{-1\downarrow}^\dagger c_{0\downarrow}^\dagger c_{1\downarrow}^\dagger c_{2\downarrow}^\dagger c_{-2\uparrow}^\dagger 
|0\rangle
\end{align}
\normalsize

\section{Master equation and Fermi's Golden Rule}

The Hamiltonian describing the Fe adatom deposited on MgO/Ag(100) coupled to a phonon bath is
\begin{align} \label{eq:SM-Hamiltonian}
H
=
\sum_i E_i C^\dagger_i C_i
+
\sum_{\eta}
\omega_\eta b^\dagger_\eta b_\eta
+
\sum_{i,f,\eta}
G_{if}^\eta \, C^\dagger_f C_i \, (b^\dagger_\eta + b_\eta)
,\end{align}
where $E_i$ are the energies of the states obtained from the Stevens Hamiltonian \eqref{eq:Fe-Stevens-Hamiltonian-Baumann}, $\omega_\eta$ are the phonon frequencies obtained from DFT calculations and $G_{if}^\eta$ are the electron-phonon matrix elements between the adatom multiplet states. $c^\dagger$ ($b^\dagger$) and $c$ ($b$) are the electron (phonon) creation and annihilation operators, respectively. The first term, $H_A$, descirbes the electronic structure of the adatom and the second one, $H_{B}$, the phonon bath. The last term, $V$, represents the interaction between the adatom electronic states and the phonon bath.

Using the density matrix formalism, the time evolution of the whole system is given by the Liouville-von-Neumann equation in the interaction picture
\begin{equation}
\dfrac{d \rho(t)}{d t}
=
-i\left[ V , \rho(t) \right]
.\end{equation}
However, as the system of interest is the adatom itself, and not the phonon bath, the problem can be simplified making use of the theory of open quantum systems~\cite{Breuer2007}. The state of the adatom system (S) can then be obtained from a partial trace over the phonon bath degrees of freedom (B), introducing the so called reduced density matrix:
\begin{align}
\rho_A(t) = \Tr_B \rho(t)
.\end{align}

In pump probe experiments preformed to measure spin-flip lifetimes, an excited state of the adatom is populated by a pump pulse, and then the evolution towards the ground states is measured with probe pulses. In the density matrix formalism, the evolution of the occupations of the electronic states of the adatom, is given by the diagonal elements of the reduced density matrix, $\rho$ from now on. Expanding the Liouville-von-Neumann equation and after several approximations, known as the Born-Markov approximations, we obtain the master equation for the diagonal elements of the density matrix of the adatom:
\begin{align}
\nonumber
\dfrac{d\rho_{ii}(t)}{d t}
=
&+
2\pi
\sum_{j,\eta}
|{G}_{ij}^\eta|^2
\rho_{jj}(t)
\left[
n_{BE}^\eta \delta(E_{j}-E_{i}+\omega_\eta)
+
(n_{BE}^\eta+1) \delta(E_{j}-E_{i}-\omega_\eta)
\right]
\\
&-
2\pi
\sum_{j,\eta}
|{G}_{ij}^\eta|^2
\rho_{ii}(t)
\left[
n_{BE}^\eta \delta(E_{i}-E_{j}+\omega_\eta)
+
(n_{BE}^\eta+1) \delta(E_{i}-E_{j}-\omega_\eta)
\right]
.\end{align}
Here $n_{BE}^\eta$ represents the thermal occupation of a phonon $\omega_\eta$ given by the Bose–Einstein distribution function.

At low temperatures it is a safe assumption to consider only the initial excited state $\Psi_i=\Psi_1$ and final ground state $\Psi_f=\Psi_0$, with $E_i-E_f=\Delta E>0$:
\begin{align}
\begin{pmatrix}
d\rho_{11}/d t
\\
d\rho_{00}/d t
\end{pmatrix}
=
\begin{pmatrix}
-(\Gamma_0+\Gamma_T)
&
+\Gamma_T
\\
+(\Gamma_0+\Gamma_T)
&
-\Gamma_T
\end{pmatrix}
\cdot
\begin{pmatrix}
\rho_{11}
\\
\rho_{00}
\end{pmatrix}
,\end{align}
where we have defined the temperature independent rate
\begin{align}
\Gamma_0
=
2\pi\sum_{\eta}|{G}_{1,0}^\eta|^2 \delta(\Delta E-\omega_\eta)
,\end{align}
and the temperature dependent rate
\begin{align}
\Gamma_T
=
2\pi\sum_{\eta}|{G}_{1,0}^\eta|^2 n_{BE}^\eta \delta(\Delta E-\omega_\eta)
.\end{align}

This set of coupled differential equations has a simple solution in the basis of eigenvectors that diagonalizes the matrix above:
\begin{align}
\begin{pmatrix}
\rho_{1 1}(t)
\\
\rho_{0 0}(t)
\end{pmatrix}
=
C_1
\begin{pmatrix}
\Gamma_T/(\Gamma_0+\Gamma_T)
\\
1
\end{pmatrix}
+
C_2
\begin{pmatrix}
1
\\
-1
\end{pmatrix}
e^{-(\Gamma_0+2\Gamma_T)t}
.\end{align}
Where $C_1$ and $C_2$ are constants to be determined by the initial conditions of the system. If the adatom is prepared tp be in the excited state $\Psi_1$, then $\rho_{11}(0)=1$ and $\rho_{00}(0)=0$, thus the evolution of the density matrix is given by
\begin{align}
\rho_{1 1}(t)
=
\frac{\Gamma_T}{\Gamma_0+2\Gamma_T}
+
\frac{\Gamma_0+\Gamma_T}{\Gamma_0+2\Gamma_T}
e^{-(\Gamma_0+2\Gamma_T)t}
,\end{align}
and
\begin{align}
\rho_{0 0}(t)
=
\frac{\Gamma_0+\Gamma_T}{\Gamma_0+2\Gamma_T}
-
\frac{\Gamma_0+\Gamma_T}{\Gamma_0+2\Gamma_T}
e^{-(\Gamma_0+2\Gamma_T)t}
.\end{align}

Thus, the Fermi's Golden Rule rate equation presented on the main text (Eq.~\eqref{eq:fermis-golden-rule}) can be inferred from these equations:
\begin{align}
\Gamma_{1\rightarrow0}
=
\Gamma_0+2\Gamma_T
=
2\pi
\sum_{\eta}
|{G}_{1,0}^\eta|^2
\left[ 2n_{BE}(\omega_\eta)+1 \right]
\delta(E_{1}-E_{0}-\omega_\eta)
.\end{align}

\putbib[./bibliografia]

\end{bibunit}

\end{document}